\documentclass[conference]{IEEEtran}
\usepackage{cite}
\usepackage{amsmath,amssymb,amsfonts}
\usepackage{algorithmic}
\usepackage{graphicx}
\usepackage{textcomp}
\usepackage{xcolor}
\usepackage{amsthm}
\usepackage{acronym}
\usepackage{url}

\usepackage{breakurl} 
\usepackage{caption}
\usepackage{subcaption}
\usepackage{multirow}
\usepackage{balance}
\usepackage{tabularray} 

\acrodef{API}{Application Program Interface}
\acrodef{TDLib}{Telegram Database Library}

\begin{document}

\theoremstyle{definition}
\newtheorem{definition}{Definition}[section]

\newcommand{\TODO}{\textcolor{blue}{\textbf{TODO}}}
\newcommand{\comm}[1]{\textbf{\color{cyan}{#1}}}
\newcommand{\alice}{$\mathcal{A}$}
\newcommand{\bob}{$\mathcal{B}$}
\newcommand{\pn}{{\em People Nearby}}

\title{Watch Nearby!\\ Privacy Analysis of the {\em People Nearby} Service of Telegram}

\author{
    \IEEEauthorblockN{
    Maurantonio Caprolu\IEEEauthorrefmark{1},
    Savio Sciancalepore\IEEEauthorrefmark{2}, 
    Aleksandar Grigorov\IEEEauthorrefmark{2}, 
    Velyan Kolev\IEEEauthorrefmark{2}, 
    Gabriele Oligeri\IEEEauthorrefmark{3}}\\
    \IEEEauthorblockA{    
    \IEEEauthorrefmark{1} King Abdullah University of Science and Technology \\
    RC3 Center, CEMSE Division -- Thuwal, Saudi Arabia\\
    maurantonio.caprolu@kaust.edu.sa\\
    \IEEEauthorrefmark{2} Eindhoven University of Technology -- Eindhoven, Netherlands\\
    s.sciancalepore@tue.nl, \{a.grigorov, v.kolev\}@student.tue.nl\\
    \IEEEauthorrefmark{3} Division of Information and Computing Technology,\\ College of Science and Engineering, Hamad Bin Khalifa University, Qatar Foundation - Doha, Qatar\\
    goligeri@hbku.edu.qa
    }
}

\maketitle


\begin{abstract}
\emph{People Nearby} is a service offered by Telegram that allows a user to discover other Telegram users, based only on geographical proximity. Nearby users are reported with a \emph{rough} estimate of their distance from the position of the reference user, allowing Telegram to claim location privacy.
In this paper, we systematically analyze the location privacy provided by Telegram to users of the \emph{People Nearby} service. Through an extensive measurement campaign run by spoofing the user's location all over the world, we reverse-engineer the algorithm adopted by \emph{People Nearby} to compute distances between users. Although the service protects against precise user localization, we demonstrate that location privacy is always lower than the one declared by Telegram ($500$~meters). 
Specifically, we discover that location privacy is a function of the geographical position of the user. Indeed, the radius of the location privacy area (localization error) spans between $400$~meters (close to the equator) and $128$~meters (close to the poles), with a difference of up to 75\% (worst case) compared to what Telegram declares. After our responsible disclosure, Telegram updated the FAQ associated with the service.
%
Finally, we provide some solutions and countermeasures that Telegram can implement to improve location privacy. In general, the reported findings highlight the significant privacy risks associated with using \emph{People Nearby} service.
\end{abstract}


\section{Introduction}
\label{sec:introduction}
Instant messaging has become an integral part of modern communication, both for personal and professional use~\cite{rosler2018_eurosp},\cite{alkhulaiwi2016_pst}. In this context, Telegram is one of the most popular instant messaging applications. It was founded in 2013 and is based on a centralized Cloud-based architecture, with Cloud servers deployed worldwide and the operational center located in Dubai, UAE~\cite{dargahi2017_cikm}. According to recent statistics, Telegram accounts for 550 million active users monthly and the average Telegram user spends approximately 3 hours per day using the related mobile cellular app~\cite{backlinko_telegram}.

Security and privacy have been among the selling points of Telegram since its first release~\cite{albrecht2022_sp}. In particular, Telegram supports the privacy of users through several means, including, e.g., encrypted end-to-end chats, device-specific communications, self-destructive messages~\cite{abu2017_eurousec}, and in recent years, it has even supported dedicated cryptography contests~\cite{contest_telegram}. Partly due to the perceived enhanced users' privacy, Telegram has been used in various controversial situations, e.g., terrorism and far-right groups, crypto investors, and for the exchange of illegal materials~\cite{morgia2023_icws},\cite{illegal_telegram}.

As a distinctive feature, Telegram offers various social-like services, such as channels, groups, and various location-based services. One of such location-based services is \pn, released to the public in June 2019~\cite{pn_telegram}~\cite{hartle2022impact}. \pn\ allows users to discover other Telegram users, without even being in their contact list and without knowing the telephone number, but only based on geographical proximity. For each user who opts in to participate in the service, Telegram reports the username, a profile photo, and a rough indication of the distance of the remote user from the current user location (see Fig.~\ref{fig:telegram_distance} in Sec.~\ref{sec:people_nearby}). Users can also create location-based groups, where they can invite anyone in their proximity and exchange messages in groups.

In recent years, the \pn\ service of Telegram has been reported in various news related to privacy. Some hobbyists~\cite{mapGithub} and news media~\cite{news_telegram} discussed the threat of triangulating the location of users based on the information displayed in the app, which could affect user privacy. To mitigate such threats, Telegram has updated the functionality multiple times and also released statements delegating privacy issues to the awareness of users who explicitly accept the terms of the app~\cite{telegram_updates}. However, to the best of our knowledge, no scientific contribution systematically investigated the actual location privacy provided by Telegram to users of the \pn\ service. Moreover, no studies provide an in-depth description of how Telegram implements mutual distance computation. Finally, although some scientific papers have already investigated scams carried out through Telegram (see Sect.~\ref{sec:related_work} for an overview), no contributions have investigated scams carried out through the \pn\ service.

{\bf Contribution.} In this paper, we carry out a systematic study on the privacy properties of the \pn\ service offered by Telegram. Specifically, we provide the following main contributions.
\begin{itemize}
    \item We reverse-engineer the methodology used by the \pn\ service to report remote users' distances from a given user location. Through an extensive real-world measurement campaign involving 302 independent measurements, we demonstrate that Telegram trades off distance accuracy with location privacy, making precise user localization impossible.
    \item We show that the actual location privacy of \pn\ is always less than $500$~meters---this one being declared by the service, and this depends on the location of the target user. In particular, while the radius of the uncertainty area at lower latitudes is about 400~meters,  the radius of the uncertainty area for users at higher latitudes decreases to about 128~meters---being 25\% of the declared value. 
    \item We discuss potential solutions and countermeasures that Telegram can implement either to improve users' location privacy or to make the information displayed to the user consistent with the actual amount of provided location privacy. 
\end{itemize}

{\bf Paper organization.} The rest of this paper is organized as follows. Sect.~\ref{sec:related_work} reviews the recent scientific literature on Telegram, Sect.~\ref{sec:people_nearby} introduces the service \emph{People Nearby} of Telegram, Sect.~\ref{sec:methodology} describes the methodology used to gather the data, Sect.~\ref{sec:measurement_campaign} describes our measurement campaign, Sect.~\ref{sec:reverse} shows how the service \emph{People Nearby} works, Sect.~\ref{sec:loc_priv_analysis} analyzes the actual degree of location privacy offered to users of \emph{People Nearby}, Sect.~\ref{sec:impact} discusses responsible disclosure and ethical aspects of our research, and, finally, Sect.~\ref{sec:conclusion} draws the conclusion. 

\section{Related work}
\label{sec:related_work}
Many free instant messaging applications are available on the market that allow people to communicate with each other through text, voice, and media. Telegram is an open-source application with high-security features~\cite{Sutikno2016WhatsAppVA}~\cite{kokevs2017}, which makes it perfect for scenarios affected by censorship~\cite{kargar2018censorship,ermoshina2021telegram} and, unfortunately, for the distribution of illicit content~\cite{guhl2020safe,walther2021us}. 

Telegram has received a lot of attention from the security research community in recent years, especially due to its declared focus on security and privacy. For instance, Ludant et al.~\cite{ludant2023_sp} identified privacy leaks in Telegram that, combined with the usage of a 5G sniffer, would allow for traffic analysis and stealthy generation of fake traffic to a target user, based only on the availability of the mobile cellular number; Albrecht et al.~\cite{albrecht2022_sp} studied the usage of symmetric cryptography in the end-to-end encryption protocol adopted by Telegram, i.e., MTProto; similarly, Lee et al.~\cite{lee2017security} provided a security analysis of the end-to-end encryption protocol adopted by Telegram; Nobari et al.~\cite{nobari2017} carried out an analysis of the messages and the connections of the accounts; Varizipour et al.\cite{vaziripour2018_usec} analyzed the attitude towards privacy of Iranian users of Telegram; Anglano et al.~\cite{anglano201731} devised a forensic analysis based on the generation of artifacts and their retention in the device storage; Abu-Salma et al.~\cite{abu2017_eurousec} disclosed several design issues of Telegram that impact security, including an unclear description of security features such as the use of encrypted chats; and Barsocchi et al.~\cite{barsocchi2020_icqict} showed an example of an architecture for location-based services compliant with the European GDPR, using Telegram as an example to implement such a service. Hartle et al.~\cite{hartle2022impact} spoofed the position of a smartphone in a warfare scenario (Ukraine), thus being able to reach Russian and Ukrainian military forces using the {\em People Nearby} Telegram service. 

Location privacy has been an increasingly significant topic of research due to the proliferation of location-based services (LBS) and the ubiquitous nature of mobile devices. The ongoing challenge is to balance the utility of location-based services with the preservation of individual privacy, especially in the face of evolving technologies and use cases. In the context of instant messaging applications and services, Li et al.~\cite{li2010} used trace-based analysis to study how real-world users share privacy-sensitive location information. They found that user privacy concerns are correlated with their age, sex, mobility, and geographic regions. Privacy threats associated with geo-social networks are discussed by Vicente et al.~\cite{vicente2011}, investigating aspects such as location, absence, co-location, and identity. Wei et al.~\cite{wei2011} introduce a privacy attack in which the adversary uses both historic movements and friendship information to estimate the user’s trajectory. The authors also proposed a solution that allows a user to upload fake locations to protect his privacy. Also, Huaxin et al.~\cite{huaxin2018} recently introduced a privacy attack that combines different mobile social networks, being able to predict demographic attributes about users.
The applicability and effectiveness of location privacy approaches in social messaging apps are analyzed by Wernke et al.~\cite{wernke2014classification}. The authors identified protection goals such as personal information, spatial information, and temporal information. They also give an overview of the basic principles and existing approaches to protect the identified privacy goals, and finally, they classify possible attacks. 

However, to the best of our knowledge, no scientific contributions have provided an in-depth analysis of the privacy associated with any location-based service already deployed in the real world, such as \pn\ offered by Telegram.

\section{People Nearby Service}
\label{sec:people_nearby}
\begin{figure}
    \centering
    \includegraphics[width=\columnwidth]{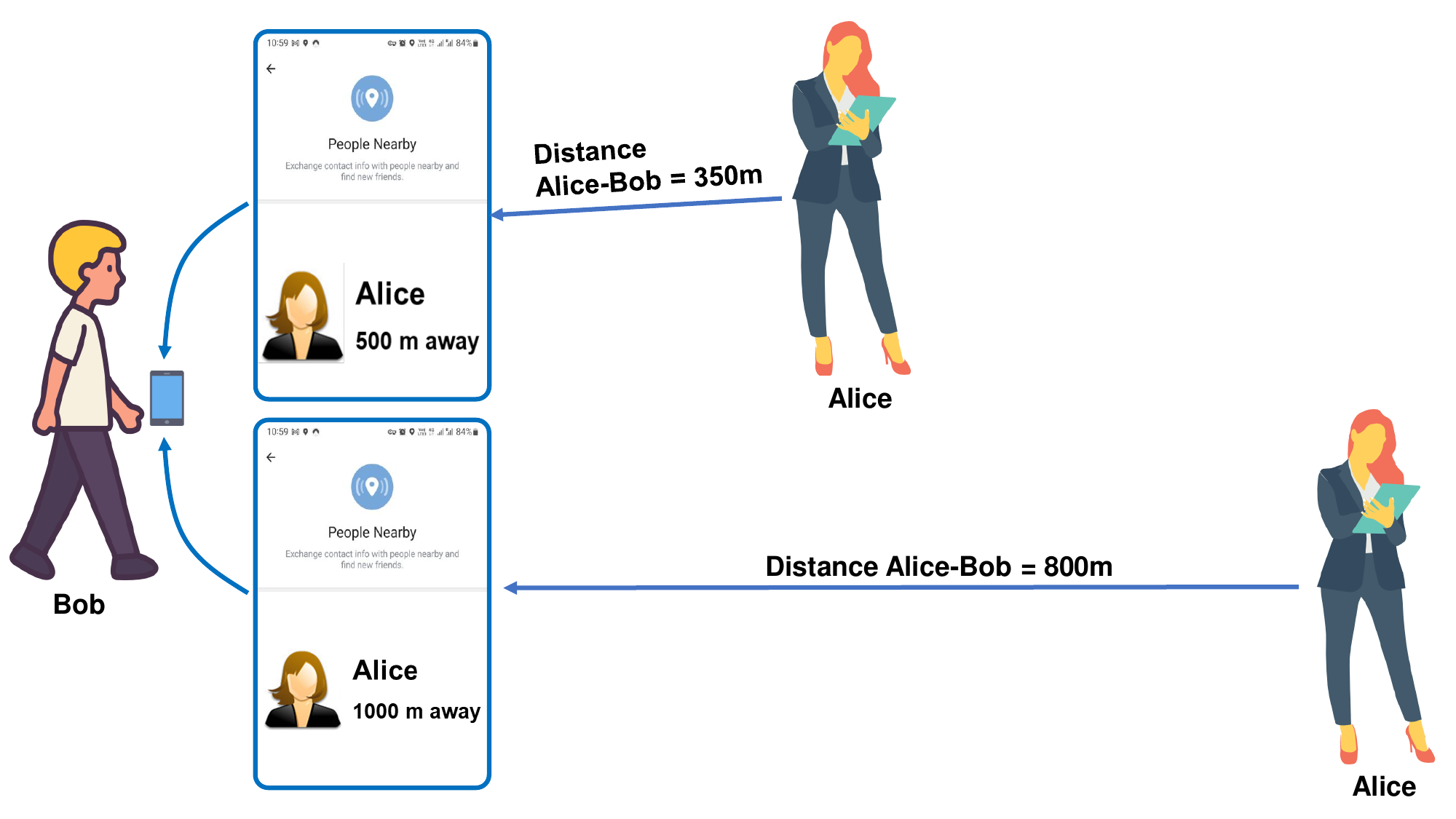}
    \caption{\pn\ service provides a rough approximation of the distance between the users in the neighborhood.} 
    \label{fig:telegram_distance}
\end{figure}

\pn\ is a service offered by Telegram that provides the user with a list of other nearby Telegram users while reporting their (rough) distance, as shown in Fig.~\ref{fig:telegram_distance}. Furthermore, \pn\ also lists location-based groups, i.e., groups associated with a particular geographical location that are visible only to nearby users. Note that location-based groups are different from Telegram \emph{super-groups}, e.g., the ones discussed in~\cite{morgia2023_icws}, which instead are not tied to any location.
Telegram provides open-source \acp{API} to allow third-party developers to create custom client applications. 
In this manuscript, we use the \ac{TDLib} library\footnote{https://core.telegram.org/tdlib}, which is a cross-platform Telegram client to build custom apps inside the Telegram platform. Specifically, we consider the {\em searchChatsNearby} API call, which returns a list of users and location-based groups nearby. 
The function takes the latitude and longitude of the user as input and returns a list containing users and groups located at different classes of distance. The returned list contains up to 100 users, while the list of groups has at most 9 entries. Each entry (being a user or a group) has an associated \emph{distance class}, chosen among one of the following: $100$~m, $500$~m, $1,000$~m, $2,000$~m, $3,000$~m, $4,000$~m, $5,000$~m, $6,000$~m, $7,000$~m, $8,000$~m, $9,000$~m, $10,000$~m, $11,000$~m, and $12,000$~m. The lists are then sorted by distance from the input location. 
Note that the returned distance class does not allow immediate accurate localization of the nearby user, and this work focuses on analyzing the location privacy provided by the service.
The information contained in the lists obtained through the {\em searchChatsNearby} function can then be used to generate further queries about specific users and groups. For this work, we investigated location privacy taking into account the distance classes of $500$~meters and $1,000$~meters. For the sake of completeness, we highlight that distances of $100$~meters are reported within the service only for users in the user's contact list, whereas higher distances, i.e., higher than or equal to $2,000$ meters, are not considered in this work, although they are reported by the service. 
%

Finally, note that our analysis was also restricted by a set of limitations related to the usage of \ac{TDLib} and the query rate of the \pn\ service. Specifically, \pn\ sets a maximum daily number of queries per user equal to $1,000$, meaning that a user cannot execute more than $1,000$~queries per day. Whenever such a limit is exceeded, \pn\ gives a temporary \emph{ban} to the user, i.e., the user cannot query the system for the next 24 hours. Also, a sudden change in the queried locations among very short consecutive time instants leads to a temporary ban. We empirically set a threshold of $90$~km/h for the maximum speed tolerated by \pn, i.e., if the change in the queried location among consecutive time instants leads to a speed estimation that exceeds such a threshold, the \pn\ service bans the user. Due to the nature of these (discovered) constraints, we believe such countermeasures have been enforced by Telegram to mitigate Denial of Service (DoS) attacks and not to protect users' location privacy. 


\section{Methodology}
\label{sec:methodology}
In the following, we provide an in-depth analysis of the location privacy provided by the Telegram \pn\ service. To carry out such an analysis, we systematically collected data on distance changes ({\em transitions}) reported by Telegram for specific accounts under our control. 

\begin{definition}
    We define a {\em transition} $(t_x)$, with $x \in [0, \ldots, \infty]$, as a change in the distance reported by the \pn\ service of Telegram from a remote user, as the result of a movement of the local or remote user. 
\end{definition}

\begin{figure}
    \centering
    \includegraphics[width=\columnwidth]{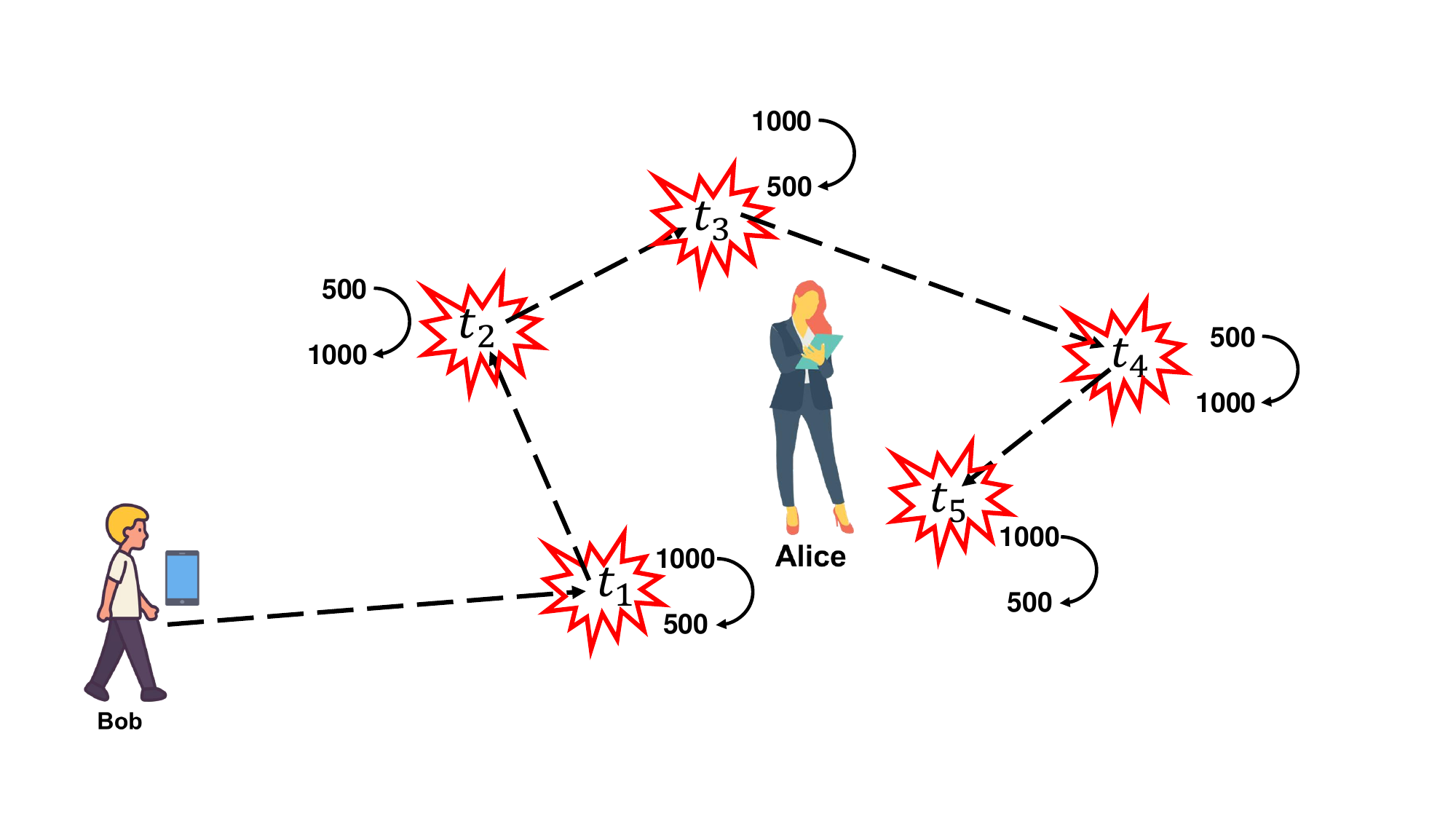}
    \caption{Data acquisition model: \bob\ repeatedly changes his trajectory when he experiences a change in the reported distance class to \alice, as shown by the \pn\ service.} 
    \label{fig:zompa_zompa_model}
\end{figure}

For our analysis, we specifically investigate transitions between $500$~meters and $1,000$~meters (and back) associated with user movements. We consider the data acquisition model in Fig.~\ref{fig:zompa_zompa_model}, with two users, \alice\ (static) and \bob\ (moving), acting as the {\em target} and the {\em finder}, respectively. 
The finder (\bob) wants to disclose \alice's position, while \alice\ wants to keep her position secret or, worst case, with the same uncertainty level claimed by the \pn\ service. In our data collection model, \bob\ walks different trajectories, repeatedly moving closer and farther from \alice. In particular, \bob\ changes his trajectory (and direction) when he experiences a change in the distance between himself and \alice\ as reported by the \pn\ service, independently of the transition being $1,000$ to $500$ or the opposite. Through this strategy, we collect a set of geographical coordinates, i.e., those where we experience a change in the distance between \alice\ (static) and \bob\ (moving).

\begin{figure}[t]
    \centering
    \includegraphics[width=\columnwidth]{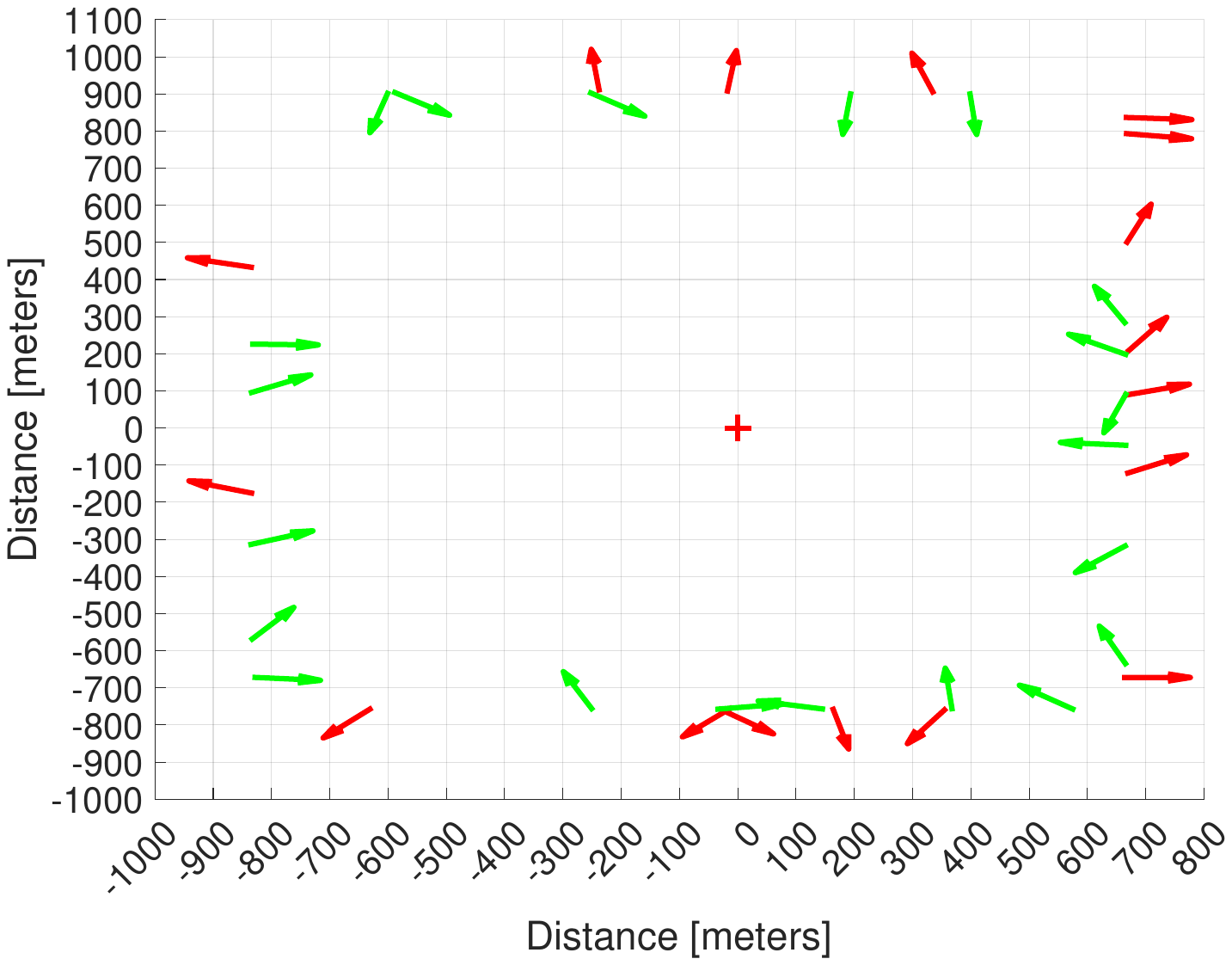}
    \caption{Collection of transitions: red arrows are related to transitions between $500$~and $1,000$~meters while green arrows report the transitions between $1,000$ and 500 meters. The tips and the tails of the arrows refer to the (two) locations---before and after---the transition is detected.} 
    \label{fig:transitions_examples}
\end{figure}

\begin{figure}
    \centering
    \includegraphics[width=\columnwidth]{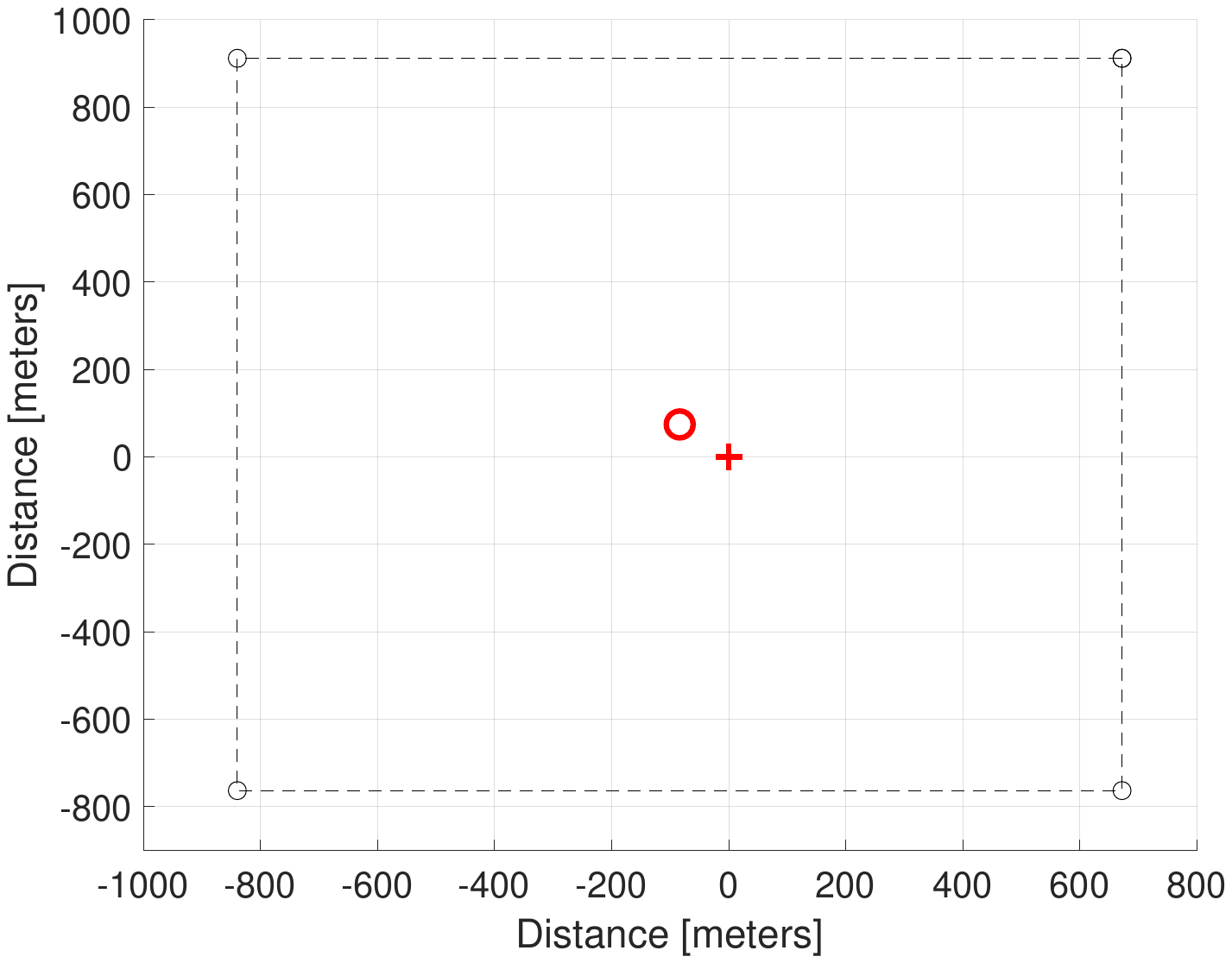}
    \caption{Modelling the transitions' distribution with a square shape: the red cross represents the position of the target, the black circles indicate the corners of the transitions' boundaries, and finally, the red circle shows the position of the centroid, being the best estimation of the target position assuming the target location is unknown to the finder.} 
    \label{fig:square_examples}
\end{figure}

Figure~\ref{fig:transitions_examples} reports the results of an exemplary real measurement. The red cross at the center $[0, 0]$~of the figure indicates the position of the target: all the geographical coordinates used in the real experiment are converted into the reference system of the target. Recalling the data collection model in Fig.~\ref{fig:zompa_zompa_model}, we performed 476 queries, collecting 38 transitions. Note that not all the queries actually collect useful transitions, since we need multiple steps to collect each single transition. The red arrows identify the transitions between $500$~and $1,000$~meters, while the green arrows denote the transitions between $1,000$~and $500$~meters. The direction of the arrows is meaningful and shows the actual trajectory of the finder. Finally, we highlight the importance of collecting the transitions in the whole surrounding of the target---this will become clear in the following sections. After discovering a transition, we compute the subsequent trajectory by setting a random direction. However, not all trajectories might go in a useful direction; thus, we empirically defined a threshold to reset the algorithm when the finder moves far away from the target.
Without loss of generality, in the following, we only consider the transitions between $500$~and $1,000$~meters (and back), i.e., both the green and red arrows in Fig.~\ref{fig:transitions_examples}. The results of Fig.~\ref{fig:transitions_examples} highlight the existence of a rectangular shape identified by the transitions, while an in-depth analysis of the transition distribution will be provided later. 
We denote by $\mathcal{T} = [t_o, \ldots, t_N]$ the set of identified transitions. We approximate their positions by considering the rectangular shape identified by the coordinates $[x_m, x_M, y_m, y_M]$, as per Eq.~\ref{eq:rectangular_shape}.
\begin{align}
  x_m & = \min_x(\mathcal{T}) \nonumber\\
  x_M & = \max_x(\mathcal{T}) \nonumber\\
  y_m & = \min_y(\mathcal{T}) \nonumber\\
  y_M & = \max_y(\mathcal{T})
  \label{eq:rectangular_shape}
\end{align}
We show the results of our analysis in Fig.~\ref{fig:square_examples}. As in the previous case (Fig.~\ref{fig:transitions_examples}), we consider all coordinates in the reference system of the target position, i.e., the red cross in the center ($[0, 0]$), while we report the edges of the rectangular shape, identified according to Eq.~\ref{eq:rectangular_shape}, with the black circles and the centroid associated with the computed edges with the red circle. Note that the centroid is the best approximation of the target position so far---assuming the position of the target is unknown to the finder.

\subsection{Data Collection} 
\label{sub:data_commection}
To implement the data acquisition model depicted in Fig.~\ref{fig:zompa_zompa_model}, we used a Samsung S10 smartphone impersonating \alice, and a DELL XPS 15 laptop impersonating \bob. The smartphone runs Android 12 and Telegram v9.6.6 (3362), while the laptop runs Ubuntu 22.04 and a home-made script to query Telegram \pn\ through the TDLib APIs.
First, we place \alice\ in a particular geographical position by spoofing the GPS of the smartphone using the app {\em Fake GPS Location Professional}\footnote{\url{https://play.google.com/store/apps/details?id=com.just4funtools.fakegpslocationprofessional&hl=en_US}}. In this way, the \pn\ service uses such a fake location as the one of \alice. Then, we designed and implemented an algorithm to emulate the movement of \bob\ around \alice, with the aim of collecting as many {\em transitions} points as possible. The algorithm iterates the following steps: (i) choosing a position for \bob, (ii) querying Telegram \pn\ from that position using the \ac{TDLib} API (specifically, the function {\em searchChatsNearby}), and (iii) retrieving \alice's distance from the returned list of \bob's nearby users. According to the position of \bob, the algorithm starts at a random position close to \alice, where the distance from \alice\ is 500 m---consequently, \bob\ is within \alice's transitions shape. Then, to choose the next position, it applies the following steps:
\begin{enumerate}
    \item \textbf{Choose the direction.} Starting from its current position, the algorithm finds a direction where it can move inside the shape while approaching its boundary. This is done by picking a random direction and checking if, after the jump, the new position remains inside the shape, i.e., the distance reported from \alice\ in \pn\ remains 500 m. The size of the jump is carefully selected not to violate the limitations in Sect.~\ref{sec:people_nearby}, to avoid our user being banned. When a suitable direction is found, the algorithm moves on to the next phase.
    \item \textbf{Probing around the shape.} The algorithm keeps jumping away from the starting point in the direction found in the previous phase. The current phase stops once the distance reported by \alice\ changes from $500$~meters to $1,000$~meters, i.e., \bob\ falls outside of the shape. Then, the next phase starts from the last identified point (the one outside the shape) and the direction that leads to it.
    \item \textbf{Finding the boundary.} In this phase, the algorithm estimates the position of the shape border (boundary) with a predetermined accuracy (in meters). Starting from the transition point identified in the previous phase, the algorithm jumps back and forth over the boundary, halving the jump distance every time to approximate the {\em transition boundary} with the desired accuracy. In all our experiments, we set such accuracy to 10 meters, to trade-off between the accuracy and the number of queries necessary to find it. 
\end{enumerate}


\section{Measurement campaign}
\label{sec:measurement_campaign}
\begin{figure}
    \centering
    \includegraphics[width=\columnwidth]{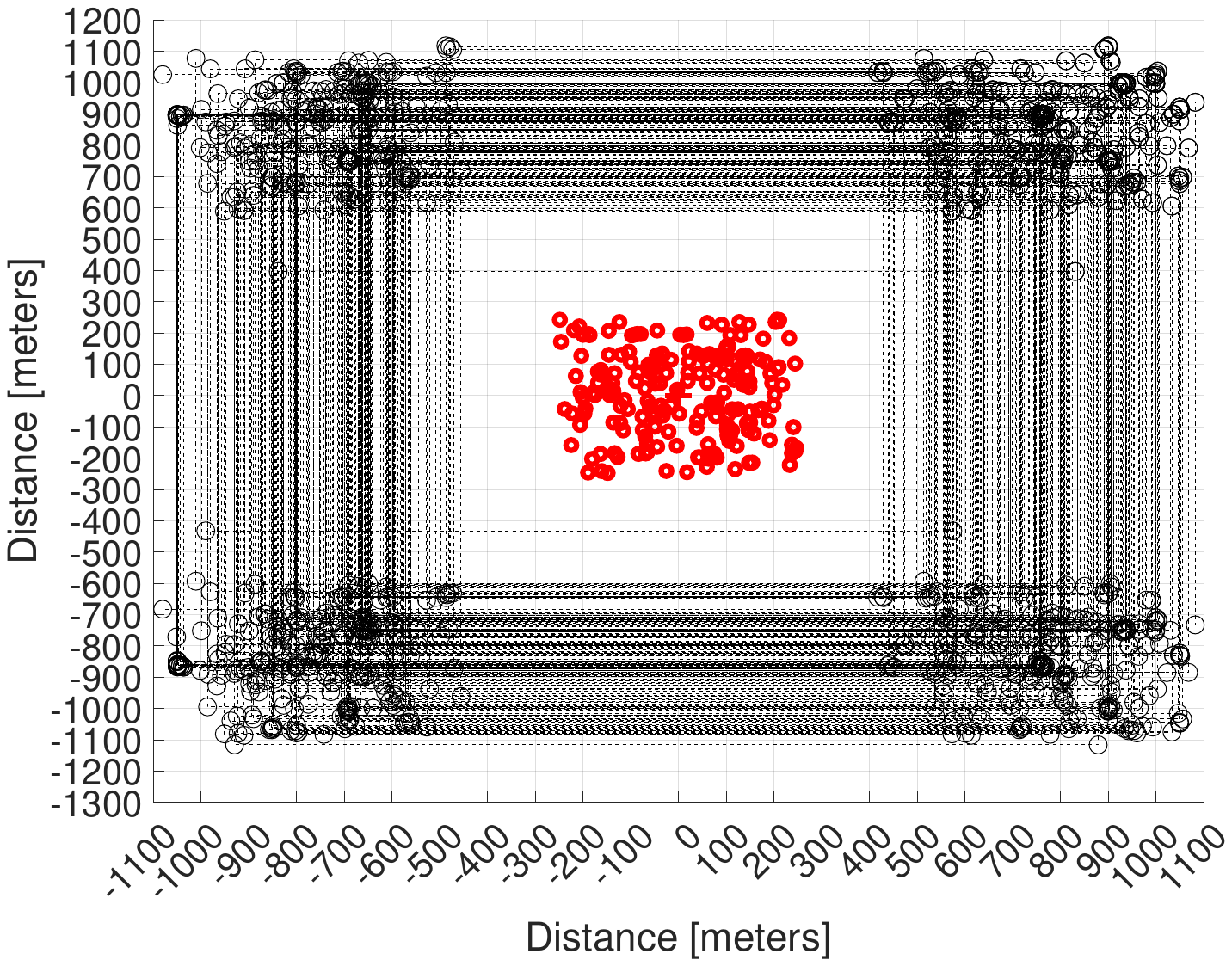}
    \caption{Measurements analysis: the black circles identify the corners of the boundaries of the transition, i.e., the transitions between the distances of $500$ and $1,000$, meters (in both the ways), red circles show the centroids, while all the targets are overlapping at $[0, 0]$.} 
    \label{fig:all_targets}
\end{figure}

We performed a total of $302$ measurements, placing a target in random positions around the world, involving the collection of a total of $8,955$ transitions. We applied the same analysis as done for Fig.~\ref{fig:square_examples}, obtaining Fig.~\ref{fig:all_targets}. All target positions (302) are placed at $[0, 0]$, and as in the previous analysis, we report the edges that approximate the position of the transitions with black circles. At the same time, we compute the red circles as the centroids (mean) of the corners of the squares. Figure~\ref{fig:all_targets} shows how the service \pn\ works to report the distance of the current user. For each target position, Telegram displaces the transitions in a pseudo-random fashion, although they are aligned in a square shape. The shape of the transitions' boundaries depends on the latitude of the target (see Sec.~\ref{sec:loc_priv_analysis}). In the following, without loss of generality, we consider only the case of square shapes. The pseudo-random offset between the square shape and the target prevents precise (meter-level) localization even when collecting many transitions in the surroundings of the target. The location privacy of the target is also shown by the distribution of the centroids (red circles), which are evenly distributed in the target's surroundings. Our subsequent analysis focuses on the distribution of the offset between the target and the transition boundaries.

\begin{figure}
    \centering
    \includegraphics[width=\columnwidth]{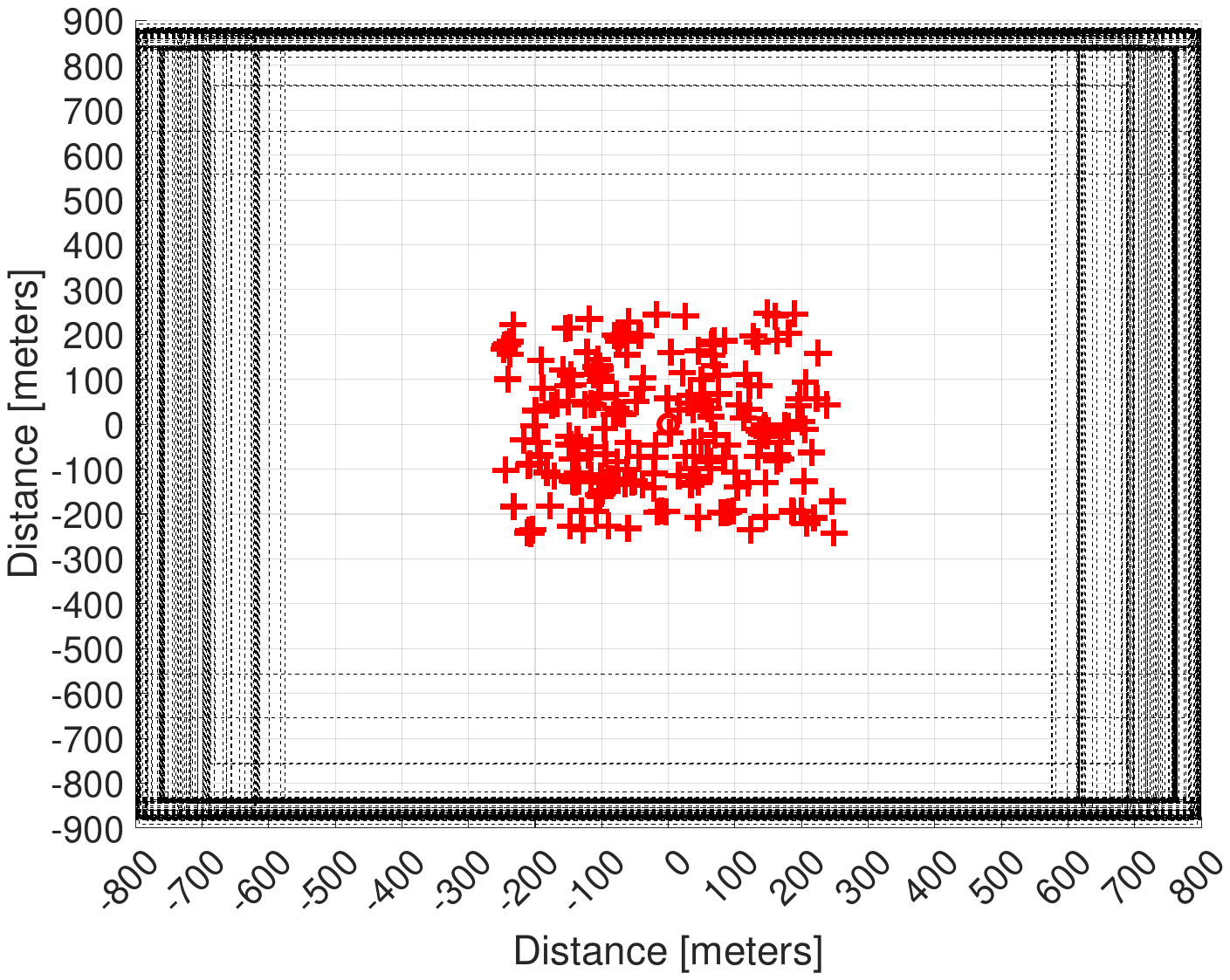}
    \caption{Measurements analysis: We overlap the centroids at the position $[0, 0]$~and the associated squares (transitions between $1,000$ and $500$ meters), while the red crosses show the actual position of the targets.} 
    \label{fig:all_centroids}
\end{figure}

We now overlap all the centroids at $[0, 0]$, as depicted in Fig.~\ref{fig:all_centroids}. Our analysis shows that the \pn\ service maps the location of the users at a pseudo-random position identified by the clouds of the red crosses---we claim the position is pseudo-random since different queries for the same account at different times at the same location return the same transitions. Moreover, note that the transition boundaries (squares) do not have the same size. In particular, we observe a larger deviation on the x-axis (longitude) than on the y-axis (latitude). A manual inspection of the measurements shows that the error might be due to the combination of multiple factors: (i) the random trajectory of our script might not have captured the most ``external" transitions as the one depicted in Fig.~\ref{fig:transitions_examples} and Fig.~\ref{fig:square_examples}, (ii) there was an out of sync between the moving spoofed position and the response to the telegram API request, or simply, an incomplete measurement is taken.

\begin{figure}[t]
    \centering
    \includegraphics[width=\columnwidth]{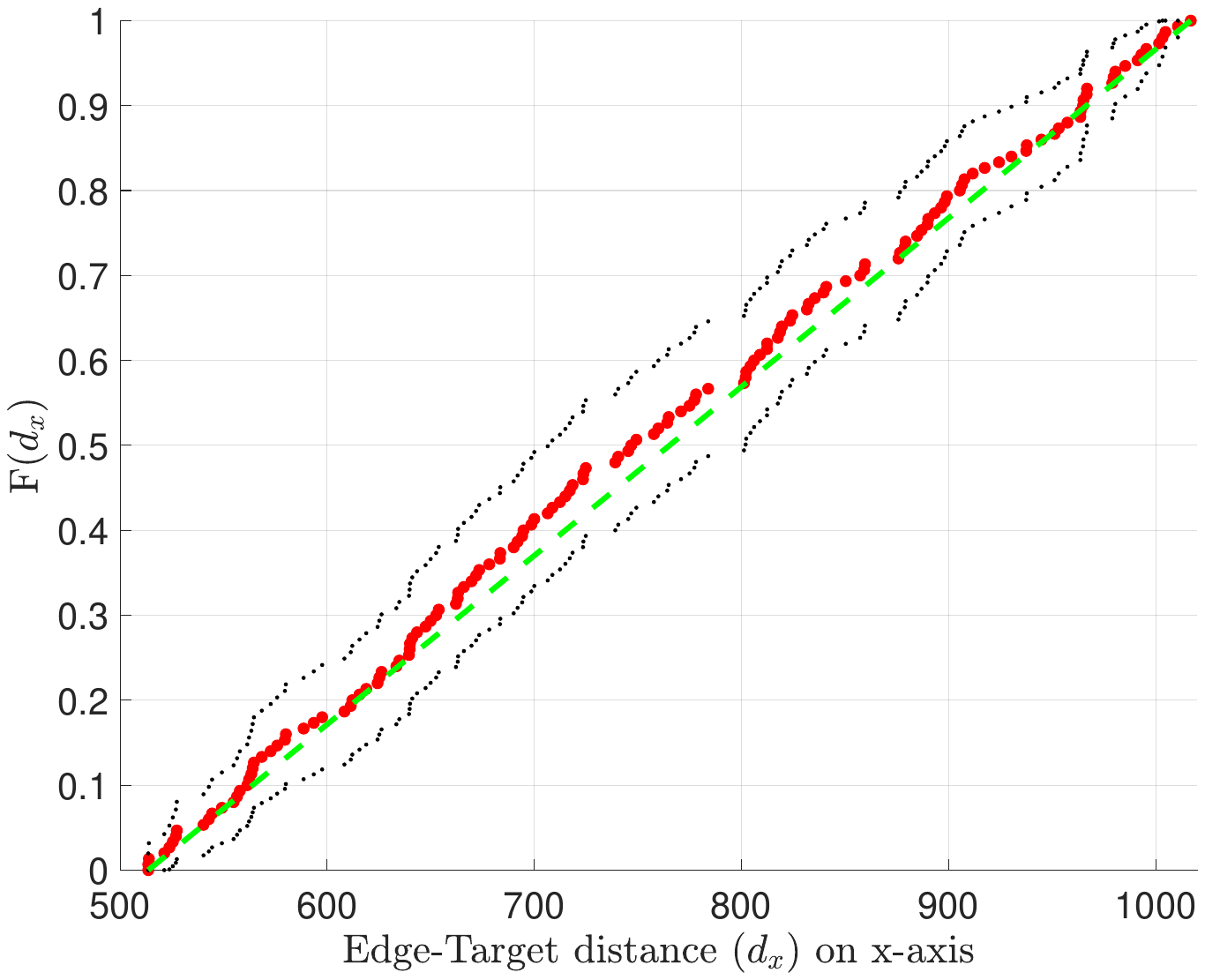}
    \caption{Target position analysis. The cumulative distribution function $F(d_x)$ is calculated on the x component of the targets. The dashed green line shows the associated uniform distribution.} 
    \label{fig:edge_target_xaxis}
\end{figure}

\begin{figure}[t]
    \centering
    \includegraphics[width=\columnwidth]{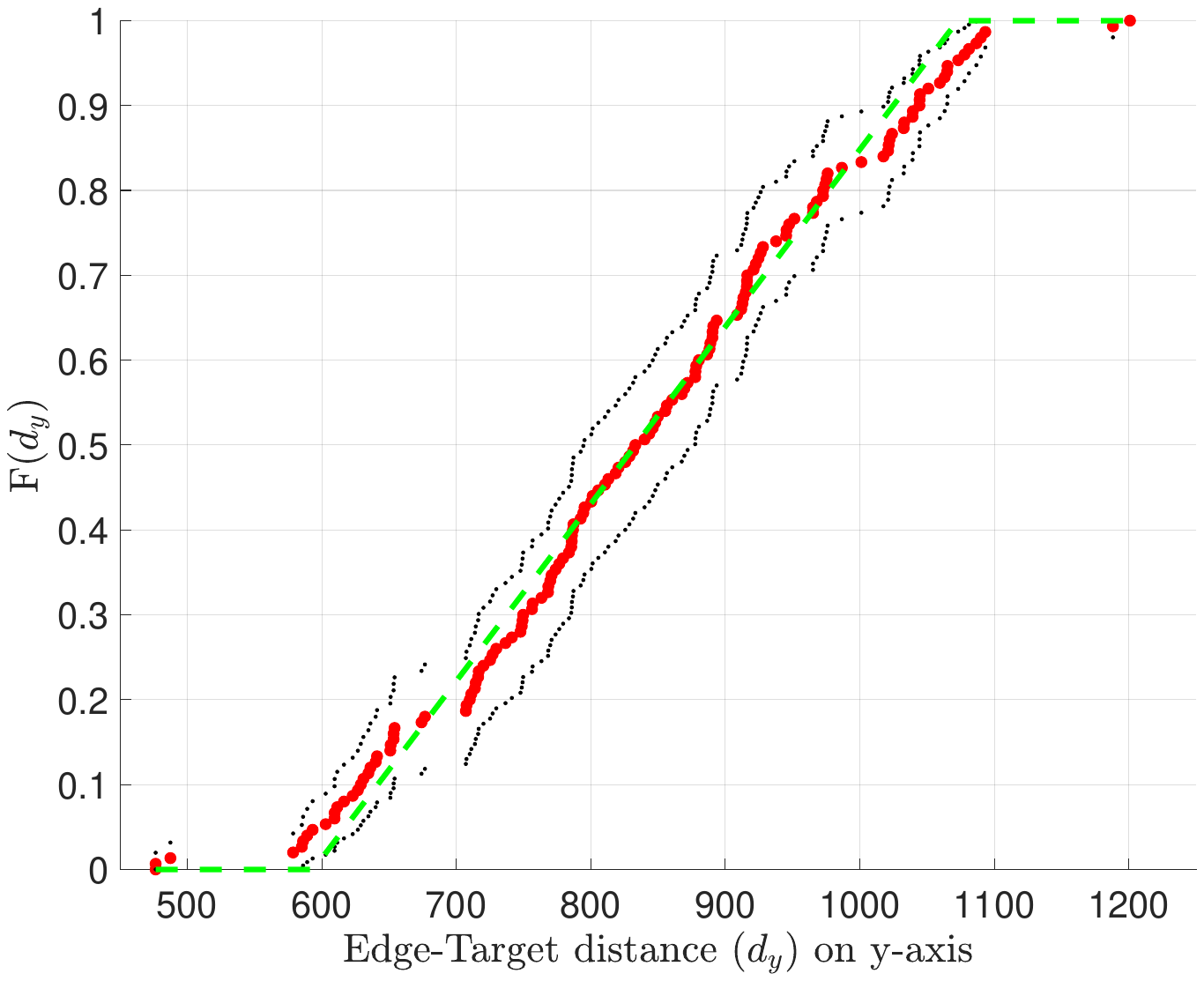}
    \caption{Target position analysis. The cumulative distribution function $F(d_y)$ is calculated on the component y of the targets. The dashed green line shows the associated uniform distribution.} 
    \label{fig:edge_target_yaxis}
\end{figure}

Given the previous considerations, we identified a set of measurements that are immune to all identified issues, and we continued our analysis. Indeed, we focus on the distribution of a subset of the targets' position in Fig.~\ref{fig:all_centroids} with respect to the edges (squares). For each pair of coordinates of the target $(x, y)$, we consider the distance between the edges and the coordinates of the target, in terms of $x$ and $y$, respectively. Figure~\ref{fig:edge_target_xaxis} shows the empirical cumulative distribution function $F(d_x)$ associated with $d_x$, i.e., the distance between the right edge of the square and the component $x$ of the target coordinate, while black dots show confidence bounds at 0.05. $d_x$ spans between about 513 meters ($F(d_x) = 0$) and about 1003 meters ($F(d_x) = 1$). We obtained similar values, i.e., about 521 and 1010 meters, considering the left edge of the square. Finally, the dashed green line shows the best fit considering a uniform distribution $\mathcal{U}_{[a, b]}$, where $a$ and $b$ are the minimum and maximum of the $x$ coordinates, respectively.
We performed the same analysis on the $y$ coordinate, as depicted in Fig.~\ref{fig:edge_target_yaxis}. Although there are some outliers, our analysis reveals that the distribution of the $y$ coordinates is also uniform, with bounds of about $[476, 1200]$.
Our analysis shows that, given a randomly deployed target, a finder can successfully collect a set of transitions, which in turn provide an upper bound on the likelihood of its position. The uncertainty limits of the location can be computed according to Fig.~\ref{fig:edge_target_xaxis} and Fig.~\ref{fig:edge_target_yaxis}, being equal to a square of about $\approx 513 \times 476$ meters. In addition, we experimentally proved that the position of the target is uniformly distributed inside that area, given the knowledge of the transitions. The approximated position of the target will be investigated in more detail in the remainder of this paper, and we will show that it can be much smaller than the one shown by Telegram, i.e., $500$ meters.

\begin{figure}%
    \centering
    \subfloat[\centering]{{\includegraphics[width=4.5cm]{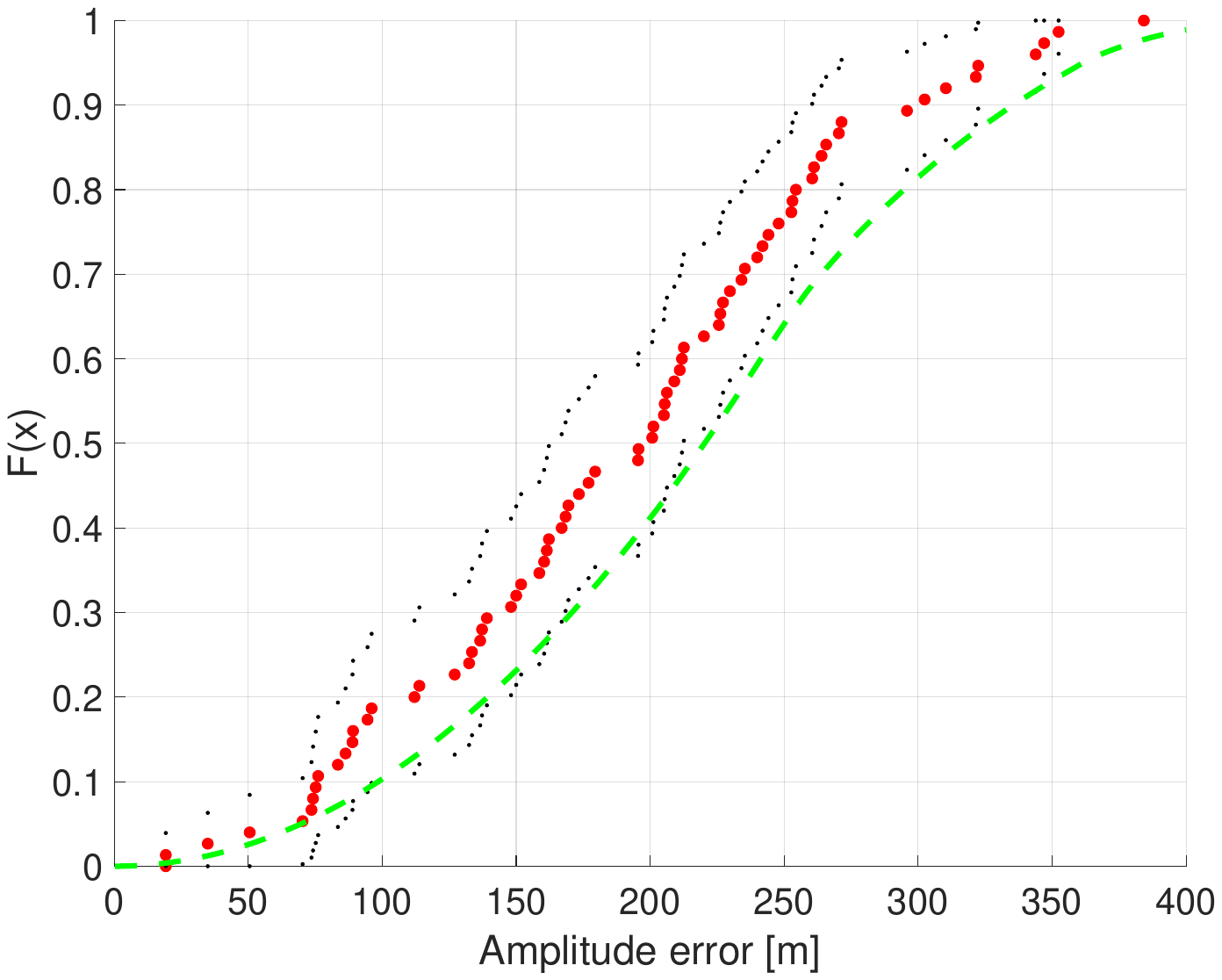} }}%
    \subfloat[\centering]{{\includegraphics[width=4.5cm]{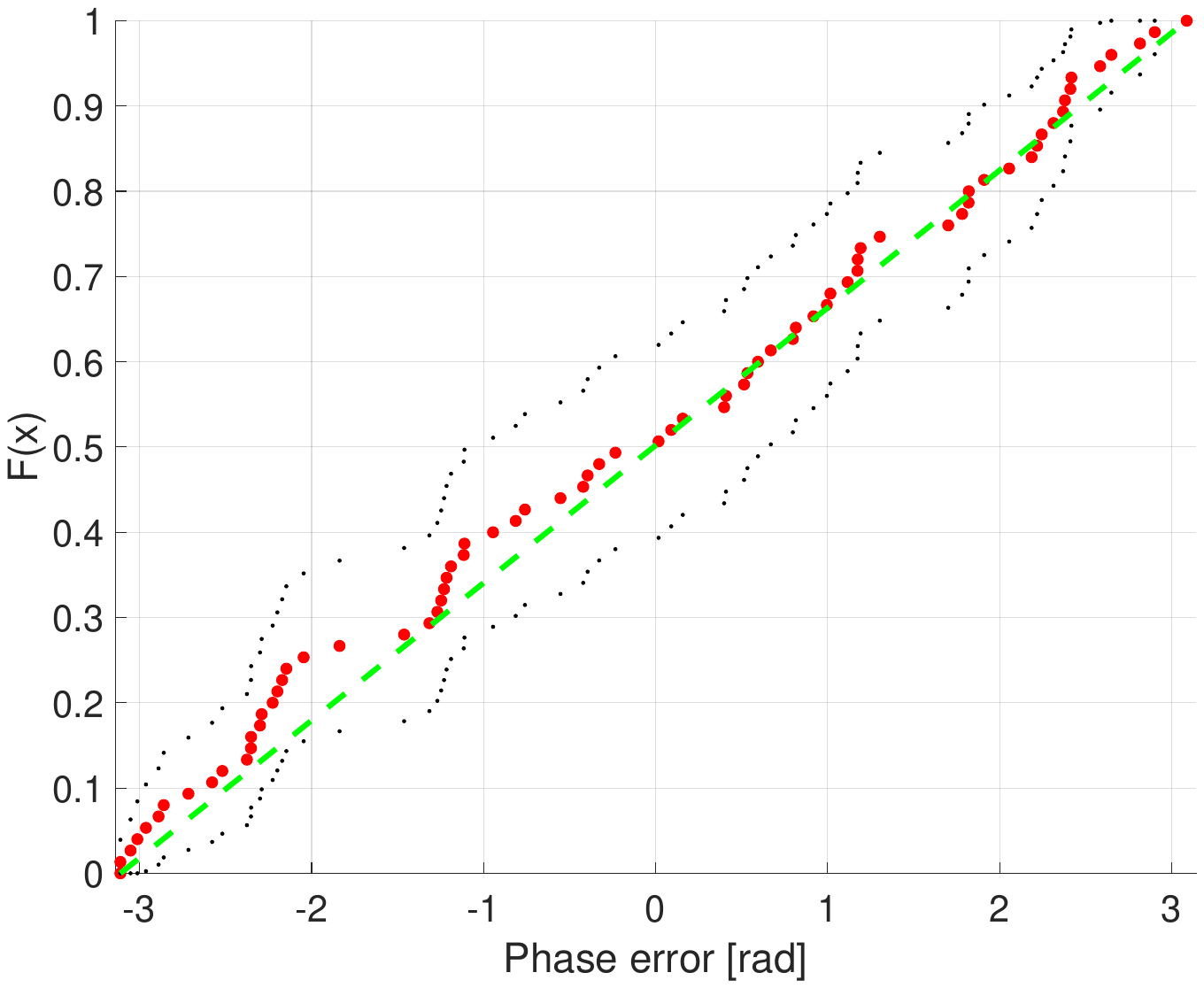} }}%
    \caption{Target localization error in terms of amplitude (a) and phase (b): the relative target position (respect to the transitions' boundaries) is uniformly distributed.}%
    \label{fig:phasor}%
\end{figure}

Finally, we consider the analysis of the amplitude and phase associated with the estimation error of the target location. Recalling Fig.~\ref{fig:all_centroids}, we want to estimate the distance between the centroids (best target position estimation) and the actual target locations. To this aim, we consider the set of phasors identified by the pairs centroid-target and, for each of them, we compute the amplitude and phase. Figure~\ref{fig:phasor} shows the results of our analysis. We confirm that the phase (Fig.~\ref{fig:phasor}(b)) is uniform in the range $[-\pi, \pi]$, as also obtained indirectly through our previous analysis. The amplitude can be estimated according to Eq.~\ref{eq:amplitude_model}, where $\rho$ is the phasor amplitude, while $x$ and $y$ are uniformly distributed random variables (as per our previous findings).
\begin{align}
    \rho = \sqrt{x^2 + y^2}
    \label{eq:amplitude_model}
\end{align}
Red dots in Fig.~\ref{fig:phasor}(a) show the empirical cumulative distribution function associated with the amplitude of the phasor. We also reported the confidence intervals (0.05) using the black dots consistent with our previous analysis. The dashed green line depicts Eq.~\ref{eq:amplitude_model} being a good fit for our empirical results. We observe that the location privacy of the target user can be arbitrarily reduced at the cost of precision. As an example, the position of the user can be estimated in a range of about 200 meters with a probability of 0.5. We stress that this analysis is cumulative with respect to all measurements. In the following, we will show that the target estimation error is a function of the target latitude. Therefore, if the adversary is aware of a rough estimation of the target position---as it is reasonable to assume---he can do much better to locate him.  

\section{Reverse engineering {\em People Nearby}}
\label{sec:reverse}
\begin{figure}
    \centering
    \includegraphics[width=\columnwidth]{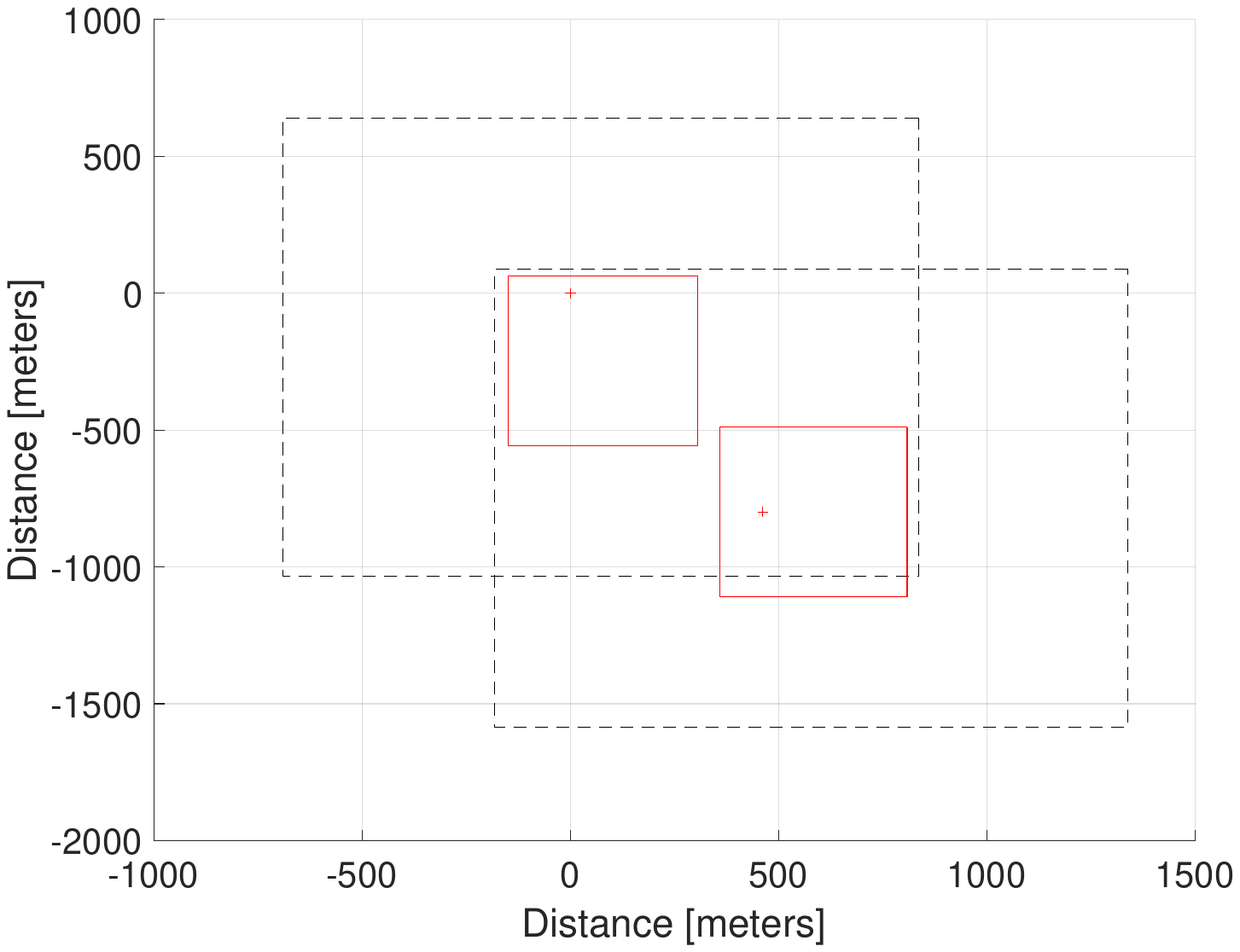}
    \caption{Transition boundaries (dashed black lines) and uncertainty regions (solid red lines) associated with 2 targets (measurements).} 
    \label{fig:reverse_eng_1}
\end{figure}

\begin{figure}
    \centering
    \includegraphics[width=\columnwidth]{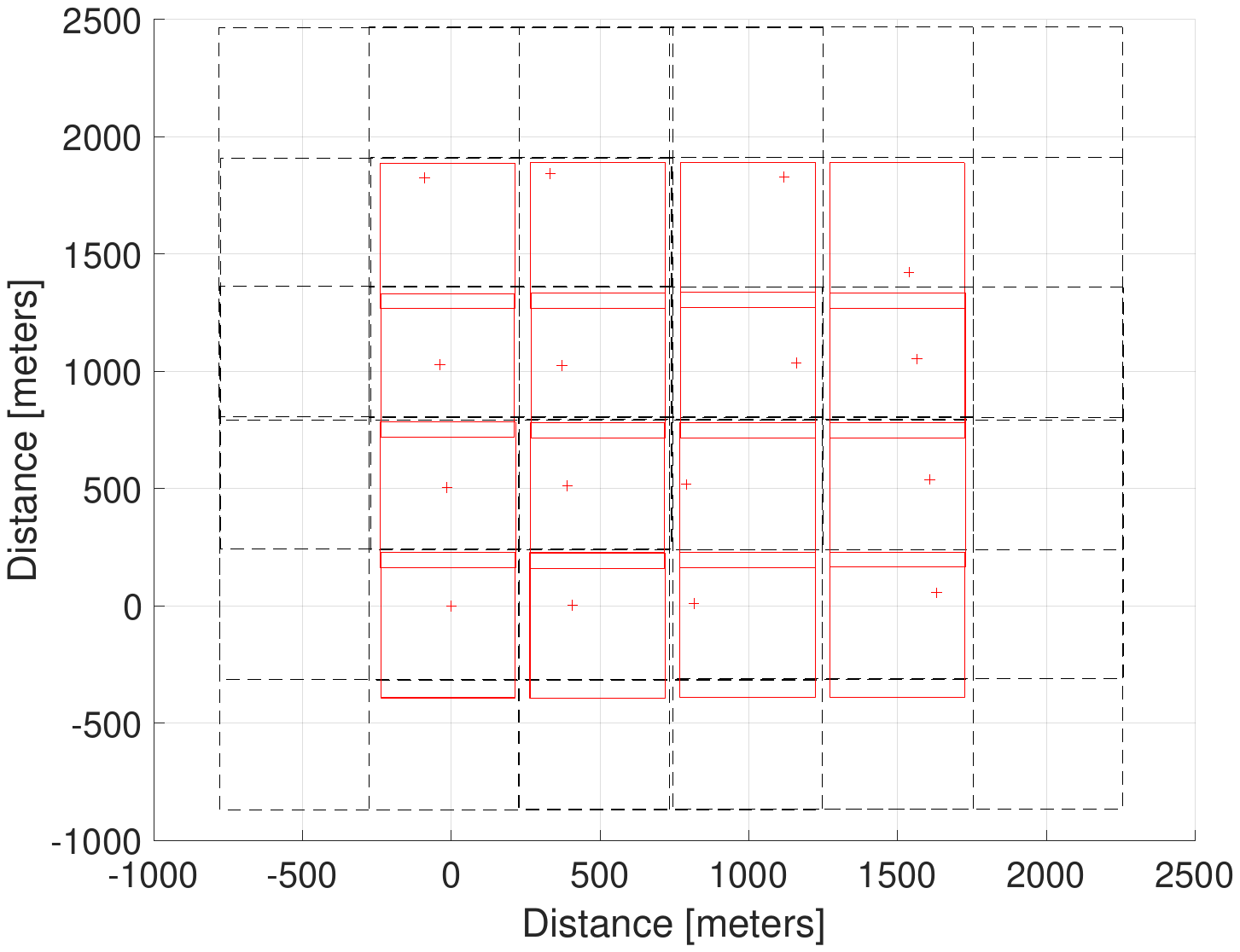}
    \caption{Transition boundaries (dashed black lines) and uncertainty regions (solid red lines) associated with 16 targets (measurements).} 
    \label{fig:reverse_eng_all}
\end{figure}

In this section, we reverse engineer how \pn\ works while providing an in-depth analysis of how it achieves location privacy. First, we recall Fig.~\ref{fig:all_centroids} and consider the boundaries identified by the distribution of the targets (red crosses). In fact, Fig.~\ref{fig:all_centroids} shows that neighbor targets (red crosses) share the same geographical distribution of the transitions (boundaries identified by the dashed black lines). As discussed in Sect.~\ref{sec:measurement_campaign}, any target belonging to the cloud represented by the red crosses cannot be uniquely identified, since Telegram generates the same transition boundaries (dashed black lines). Therefore, we compute the boundaries of the {\em uncertainty region} from all the collected measurements. Figure~\ref{fig:reverse_eng_1} shows the boundaries associated with both the transitions and the uncertainty regions considering two measurements. All coordinates have been normalized with respect to the position of the target of one of the two measurements. We stress that the black-dashed lines represent the actual boundaries of 2 measurements, while the red lines are the boundaries computed from all the measurements and expressed as their relative distance to the transition boundaries. The toy example of Fig.~\ref{fig:reverse_eng_1} highlights how the two targets belong to different uncertainty regions, and these regions are adjacent, i.e., non-overlapping on the map. 
We follow up our analysis by systematically deploying the target on neighbor positions, thus obtaining Fig.~\ref{fig:reverse_eng_all}. We considered a total of 16 measurements (target deployments depicted in Fig.~\ref{fig:reverse_eng_all} through red crosses). As considered before, for each measurement (target deployment), we report the transition boundaries (black dashed lines) and the uncertainty regions (solid red lines). We observe that the partitioning of the map has a strong symmetry for the transition boundaries and uncertainty regions. In fact, we believe that the (minor) vertical overlap of the uncertainty regions is due to the errors in the measurements and the limited number of collected measurements. We believe that location privacy is implemented through a tessellation of the playground: the map is divided into adjacent non-overlapping uncertainty regions, and the service returns the associated transition boundaries. Finally, our analysis shows that the transition boundary region has a size of about $1,500 \times 1,600$ meters, while the uncertainty region is about $500 \times 500$ meters. It is worth noting that the actual uncertainty region is about one-ninth ($\frac{1}{9}$) of the region identified by the transition boundaries. We stress that these findings are consistent with our previous analysis in Fig.~\ref{fig:all_targets} and Fig.~\ref{fig:all_centroids}.

\section{Location Privacy Analysis}
\label{sec:loc_priv_analysis}
Our final analysis focuses on the actual location privacy provided by the \pn\ service. In the following, we compare the upper bound provided by Telegram (recall Sect.~\ref{sec:people_nearby}) with the one that a malicious user can compute by leveraging the Telegram APIs. Indeed, our previous analysis (recall Sect.~\ref{sec:reverse}) shows that it is possible to estimate the uncertainty region associated with the user's position; thus, in the following, we compute the upper bound of the localization error. Assuming an uncertainty region of size $l$ meters and the target user in the center of such region, the maximum localization error (upper bound) can be calculated as $\mathcal{D} = \frac{l}{2} \sqrt{2}$ meters. As an example, we recall Fig.~\ref{fig:reverse_eng_1} and~\ref{fig:reverse_eng_all}. Approximating the square size with $l \approx 500$ meters, we have $\mathcal{D} \approx 354$ meters.

\begin{figure}[t]
    \centering
    \includegraphics[width=\columnwidth]{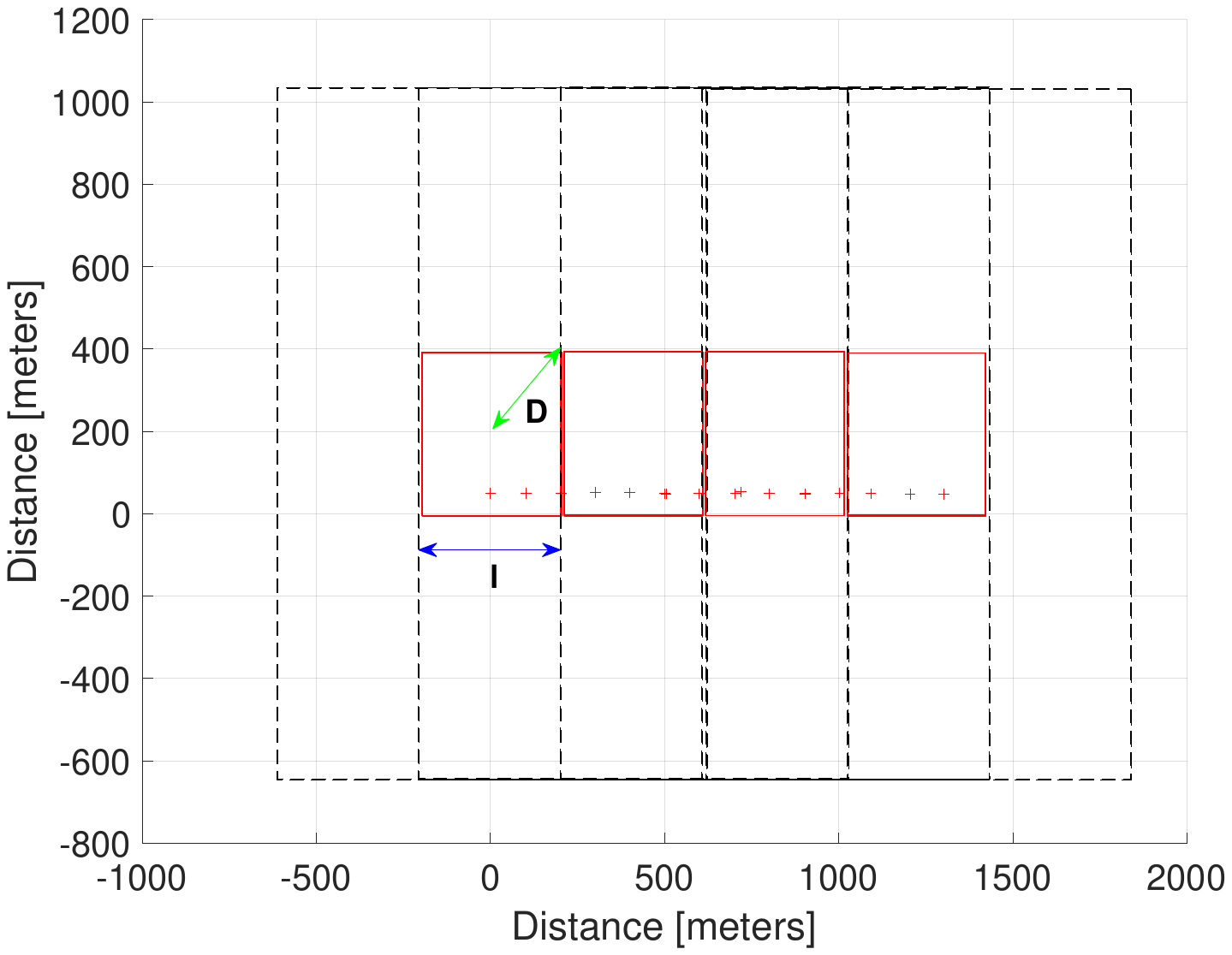}
    \caption{Methodology to compute the upper bound $\mathcal{D}$ associated with the localization error. We estimated the size of the uncertainty region $l$, and subsequently, we computed $\mathcal{D}$.} 
    \label{fig:methodology_upperbound}
\end{figure}

In the following analysis, we consider a set of (target) locations at different latitudes, as shown in Table~\ref{tab:locations}. It is worth mentioning that the calculation of $\mathcal{D}$ for each target location requires multiple measurements.  
\begin{table}
\footnotesize
\centering
\caption{\label{tab:locations}Locations considered for the estimation of the upper bound $\mathcal{D}$ associated with the localization error.}
\begin{tblr}{
  hline{2} = {-}{},
}
\textbf{City} & \textbf{Country} & \textbf{Latitude} & \textbf{Longitude} \\
Kourou        & French Guiana    & 5.154237          & -52.648526         \\
Coban         & Guatemala        & 15.46463          & -90.403683         \\
Doha          & Qatar            & 25.26174          & 51.359269          \\
Lakatamya     & Cyprus           & 35.11438          & 33.296804          \\
Carcassonne   & France           & 43.21324          & 2.344961           \\
Winnipeg      & Canada           & 50.00542          & -97.16734          \\
Malmo         & Sweden           & 55.555275         & 13.015577          \\
Helsinki      & Finland          & 60.239339         & 24.922424          \\
Bodo          & Norway           & 67.277398         & 14.374172          \\
Utqiagvik     & Alaska           & 71.300602         & -156.754113        
\end{tblr}
\end{table}
Figure~\ref{fig:methodology_upperbound} shows the methodology we adopted for the estimation of the upper bound $\mathcal{D}$ as a function of the uncertainty region size $l$. We deployed the same target at different adjacent positions, located $100$ meters from each other (as depicted by the red crosses in Fig.~\ref{fig:methodology_upperbound}). Depending on the size of $l$, multiple measurements may be required to experience a change of the transition boundaries (dashed line) and, therefore, of the uncertainty region (solid red line).
When the transition boundary shifts, we consider the size of the shift as the size of the uncertainty region $l$, thus $\mathcal{D}$ can be computed accordingly. In the (real) example of Fig.~\ref{fig:methodology_upperbound}, $l$ turns out to be about $400$ meters, while $\mathcal{D} \approx 282$ meters. Finally, we stress that experiencing 2 shifts may not be enough because of the granularity of the movements of the target, and we also verified the consistency of our estimate of $\mathcal{D}$ over multiple shifts, e.g., 4 shifts in Fig.~\ref{fig:methodology_upperbound}.

\begin{figure}
    \centering
    \includegraphics[width=\columnwidth]{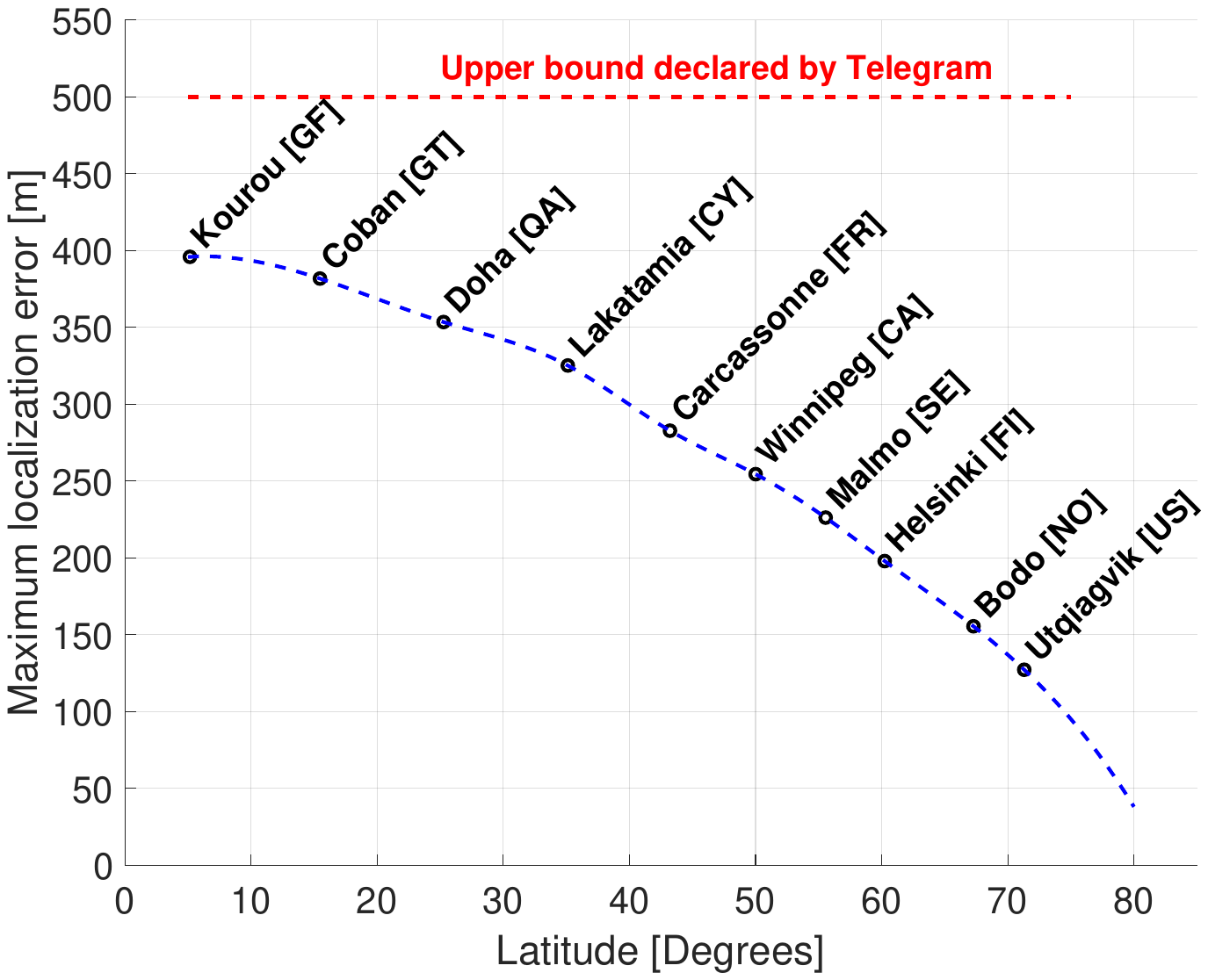}
    \caption{Maximum localization error of the target as a function of its latitude.} 
    \label{fig:privacy_latitude}
\end{figure}

We applied the same methodology to different cities at different latitudes around the world, as reported in Table~\ref{tab:locations}. Figure~\ref{fig:privacy_latitude} shows the maximum localization error $\mathcal{D}$ as a function of the latitude considering the cities in Table~\ref{tab:locations}. We observe that the upper bound on the localization error $\mathcal{D}$ is significantly affected by the target latitude; indeed, it spans between approximately $400$ meters at low latitudes (close to the Equator) and about $128$ meters at higher altitudes (close to the North Pole). It is worth noting that such distances are about 25\% and 75\% smaller than what is declared by Telegram, i.e., $500$~meters independently of the position of the target, shown by the red dashed line in Fig.~\ref{fig:privacy_latitude} (recall Sect.~\ref{sec:people_nearby}). We observe that such a value ($500$~meters) is not experienced in any of the considered locations, while being $100$~meters more than the largest distance experienced, i.e., Kourou (French Guiana), with $\mathcal{D} \approx 400$ meters.
We believe that the upper bound associated with the localization error depends on the latitude due to the tiling process performed by the Telegram algorithm (as described in Fig.~\ref{fig:reverse_eng_all}). Our intuition is that the edges of the tiles are linked to the meridians, thus implying tiles with smaller areas when the latitude increases, i.e., closer to the poles. Although our analysis did not take into account locations in the southern hemisphere, we highlight how the position of the target user could become arbitrarily small when moving north, thus significantly affecting the location privacy of the user, since the uncertainty region is indeed much smaller than the one reported to the user by Telegram.

\subsection{Shapes of transition boundaries}
\begin{figure}
\centering
\begin{minipage}{.5\columnwidth}
  \centering
  \includegraphics[width=\linewidth]{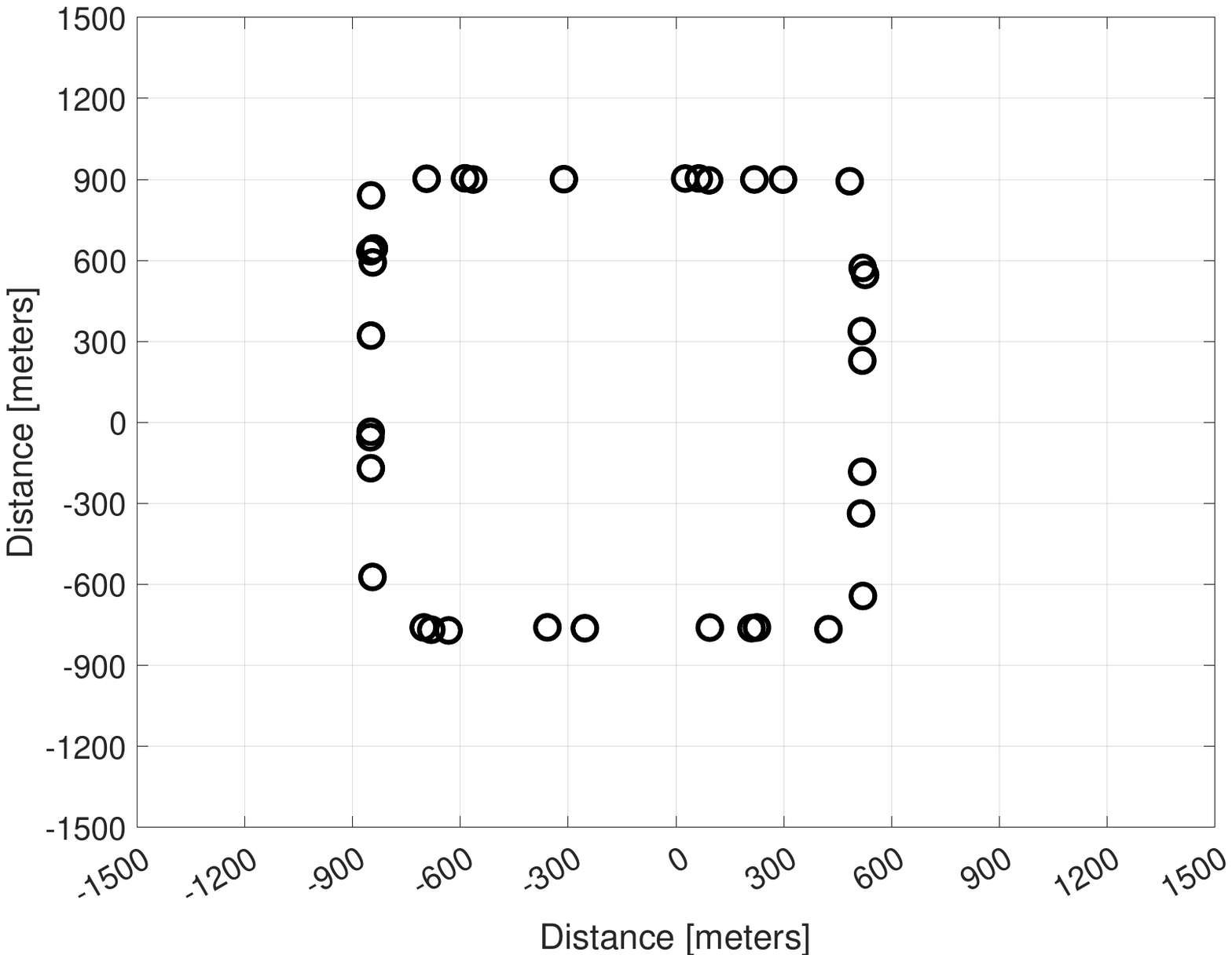}
  \caption*{(a)}
  \label{fig:test1}
\end{minipage}%
\begin{minipage}{.5\columnwidth}
  \centering
  \includegraphics[width=\linewidth]{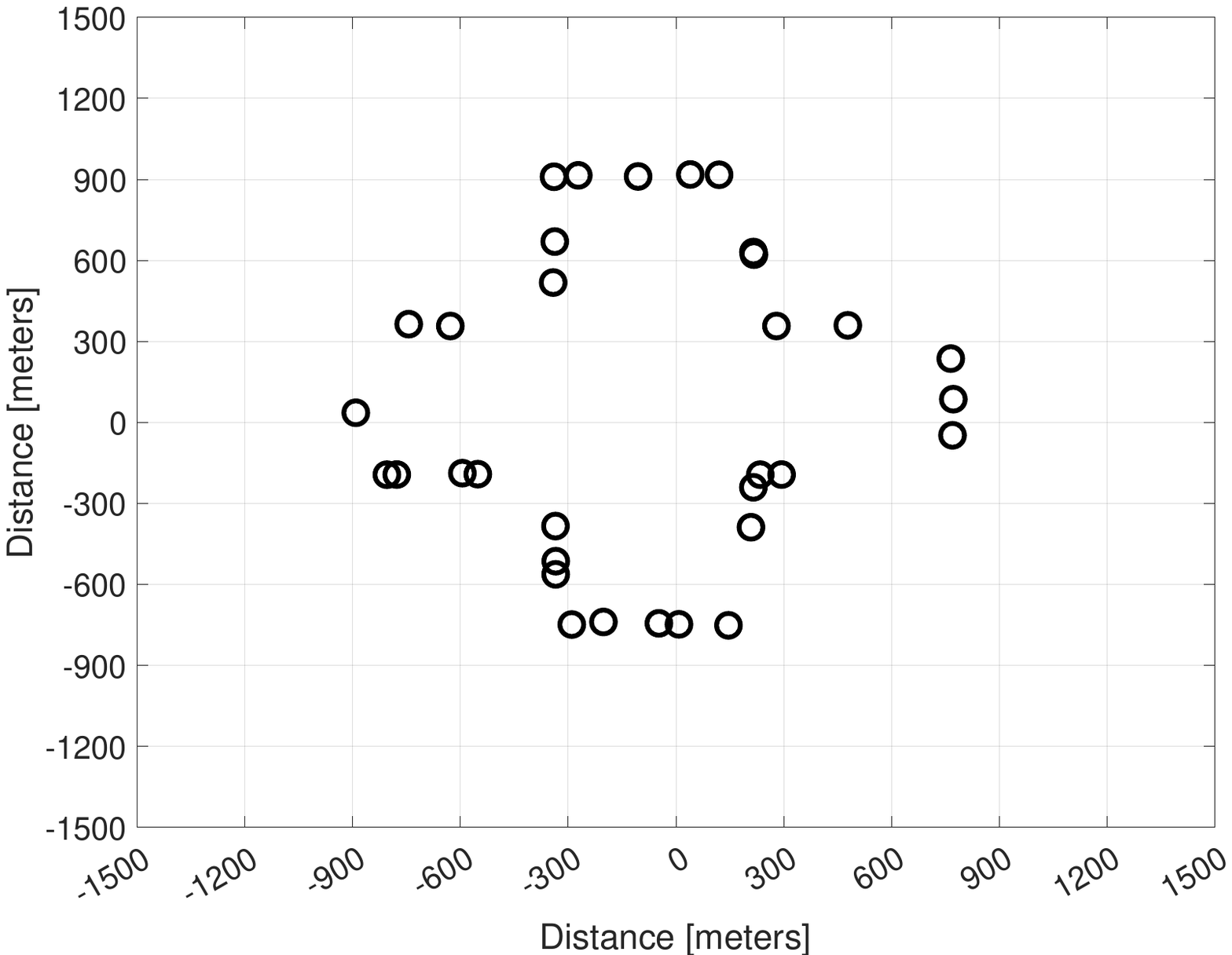}
  \caption*{(b)}
  \label{fig:test2}
\end{minipage}
\caption{Shapes of the transition boundaries: Our measurement campaign exposes two different shapes for the uncertainty regions, i.e., a square (a) and a cross shape (b).}
\label{fig:uncertainty_shape}
\end{figure}

In this section, we provide an in-depth analysis of the shapes associated with the transition boundaries as a function of the target location. Figure~\ref{fig:uncertainty_shape} shows two examples of transition boundaries collected during our measurement campaign. The square shape, i.e., Fig.~\ref{fig:uncertainty_shape}(a), has been previously discussed. Furthermore, as a function of the target position, we encountered a different shape, i.e., a cross like the one in Fig.~\ref{fig:uncertainty_shape}(b). We observe that the two shapes have similar dimensions, i.e., $1,500 \times 1,800$ meters, while for both cases the analysis of the uncertainty region is consistent with the findings reported in Sect.~\ref{sec:reverse}. 

In the subsequent analysis, we considered all measurements collected around the world and manually analyzed them to judge if the shape reassembles either a square or a cross. Figure~\ref{fig:uncertainty_shape} shows the results of our analysis, where we draw a blue square or a red cross at the same position of the target as a function of the shape associated with the transition boundary. Our findings show that the shape of the transition boundary is a function of latitude. Indeed, the world map appears split into stripes, where crosses interleave with squares. On the one hand, we stress that the shape does not affect our previous analysis (Sect.~\ref{sec:loc_priv_analysis}). Indeed, the size of the uncertainty region is the same, independently of the transition boundary shape being a square or a cross. On the other hand, we acknowledge that we cannot provide a full justification for the phenomenon.
\begin{figure}
    \centering
    \includegraphics[width=\columnwidth]{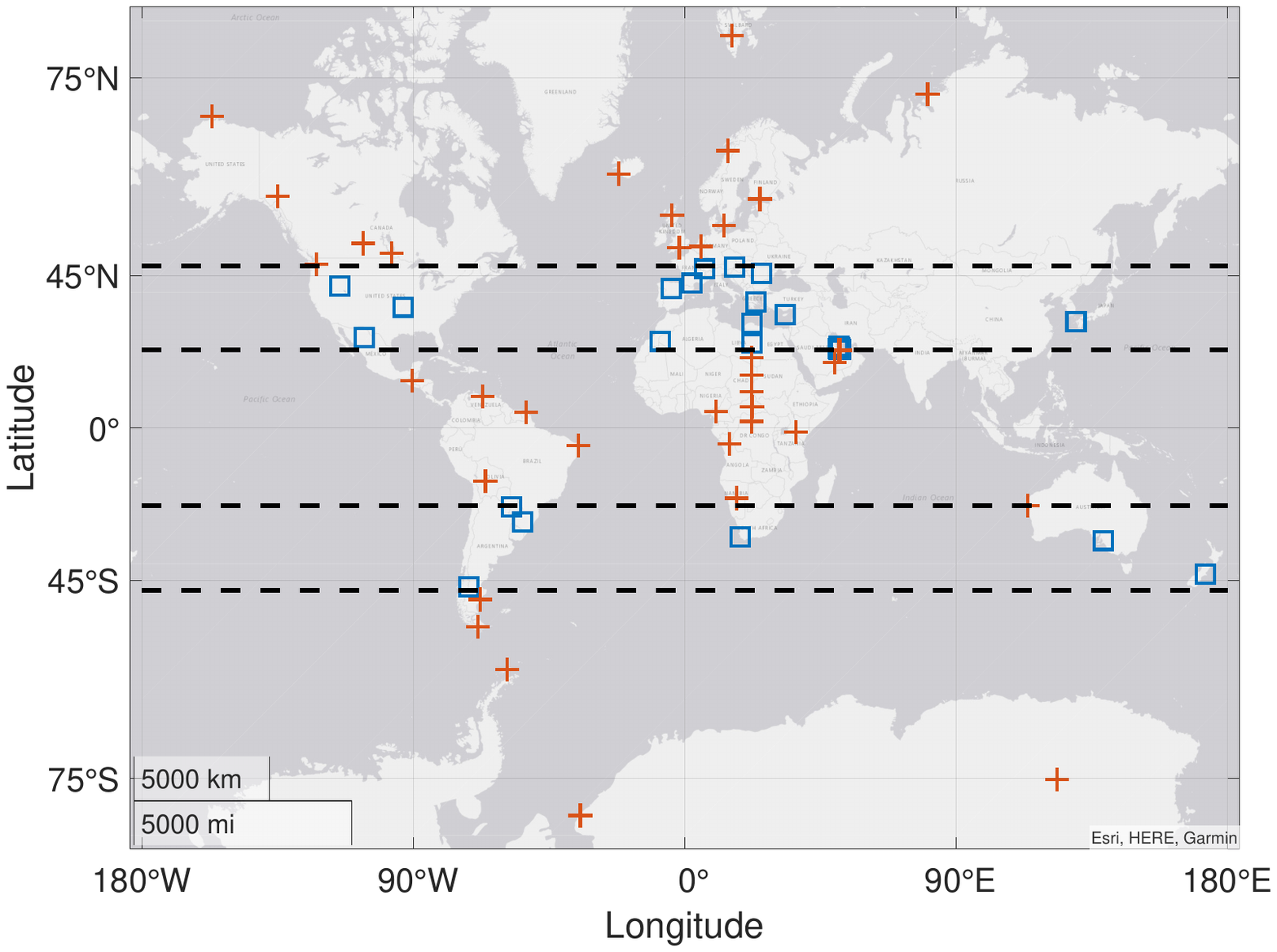}
    \caption{Shape of the transitions boundary: Square and cross patterns as a function of the target's latitude and longitude.} 
    \label{fig:geo_map}
\end{figure}

\section{Impact and Responsible Disclosure}
\label{sec:impact}
{\bf Impact of our findings.} Our analysis shows that users who activate the service \pn\ experience a location privacy much smaller than the one declared by Telegram, i.e., from 25\% (best case) to 75\% (worst case) smaller according to the user's latitude. 



{\bf Solutions and countermeasures.} 
We propose solutions and countermeasures to make users fully aware of the location privacy provided by the \pn\ service, meet the privacy level declared by Telegram, or make the attack presented in this paper more complicated (or even impossible).

\textit{Update the location privacy.} A solution involves displaying the actual location uncertainty as the distance reported for each account in the \pn\ service, rather than 500 meters (currently displayed). In this way, users can immediately identify what is the current degree of location privacy, and possibly realize that such location uncertainty changes with the current latitude value, as found in our analysis.
Another solution consists of displaying, for each user nearby, the smallest location privacy value worldwide, i.e., approximately 128 meters, rather than the 500 meters currently displayed. This is a worst-case approach, declaring to the user the lower bound of location privacy that can be guaranteed worldwide.

\textit{Increase location privacy.} This countermeasure would require a significant modification of the \pn\ service and would consist of adopting tiles with areas as a function of the current latitude, i.e., larger tiles at higher latitude values, to effectively provide an uncertainty region of 500 meters to any user, regardless of their current location. Such a modification could require a large system update of the \pn\ service, but would contribute to guaranteeing the extent of location privacy promised to users.

\textit{Change API policies.} The attack described in this paper relies on one client, i.e., the finder, moving around another user, i.e., the target, in a relatively short time frame. To mitigate this threat, we propose restricting service usage to the first geographic location declared by the user in a given time.
The API could be configured to answer user's requests within a limited area, e.g., 100 square meters, while dropping requests from more distant locations. This modification allows legitimate users to move within a limited area, such as their homes or a public square, while using the service. The allowed area can be updated after a relatively short time, for example every 10 minutes, to allow users to move consistently with the intended use of the service. On the contrary, quickly transitioning between positions 500 meters apart, as our algorithm does, would become unfeasible.
Implementing this countermeasure is simpler than the previous one, because it primarily involves request filtering. The service's fundamental logic remains intact, while the attack vector is effectively neutralized. In addition, this countermeasure will not affect legitimate clients, as the service is not intended for users on the move.


{\bf Responsible disclosure.} Transparency and ethics are of paramount importance to the authors. As such, before disclosing to the public our findings, we practiced the principle of responsible disclosure to ensure the safety and integrity of the affected parties. 
When identifying the lack of privacy associated with the service \pn\ discussed in this research, we reached out to Telegram. This was done to provide them with a comprehensive understanding of the issue and the time to take the necessary corrective measures or implement preventive strategies. The purpose was to minimize the possible harm or misuse of the information presented in our study. 

Telegram acknowledged our methodology by highlighting that \emph{``the coordinates of all points are always rounded... It is approximately 556 meters in length and width (or 787 meters diagonally) when close to equator, which is equal to the 400-meter radius result from your research''}. Moreover, they also pointed out that \emph{``we'll shortly update the FAQ with the relevant information that latitude has a minor effect on these distances.''}
Note that the decision to make our findings public, after the designated time frame, was driven by the commitment to advance knowledge in the field and to ensure that the general community is informed. We firmly believe that transparency, when combined with responsibility, can drive positive change, fostering a safer and more secure digital environment.

\section{Conclusion}
\label{sec:conclusion}
We have conducted a systematic analysis of the privacy of users of the \pn\ service featured by Telegram. We reverse-engineered the algorithm used by the \pn\ service to display rough distances between users, and experimentally showed that the actual location privacy is always less than $500$~ meters, which is the one reported by the service. Moreover, location privacy also decreases while increasing the geographical latitude of users. 
In particular, while the radius of the uncertainty area declared by Telegram is $500$~meters, our analysis shows that such a radius spans between $400$~meters and $128$~meters when the user location is close to the equator and the north pole, respectively. It should be noted that the actual uncertainty region is characterized by a radius that is approximately 25\% and 75\% smaller than that declared by Telegram as a function of the user's location.
%
%
We believe that the lack of location privacy generates significant risks for the involved users. Such concerns might motivate users to opt out of using location-based services while calling for further efforts by Telegram to protect users' privacy.

\newpage

\bibliographystyle{plain}
\balance
\bibliography{main}

\begin{thebibliography}{10}

\bibitem{contest_telegram}
{\$300,000 for Cracking Telegram Encryption}.
\newblock
  \url{https://web.archive.org/web/20190518221133/https://telegram.org/blog/cryptocontest},
  2019.
\newblock Accessed: 17-Oct-2023.

\bibitem{pn_telegram}
{Telegram adds location-flavored extras and full group ownership transfers}.
\newblock \url{https://tinyurl.com/3hv42cn4}, 2019.
\newblock Accessed: 17-Oct-2023.

\bibitem{telegram_updates}
{Telegram's People Nearby feature reveals exact user locations through
  triangulation}.
\newblock
  \url{https://www.androidpolice.com/2021/01/05/telegrams-people-nearby-feature-reveals-exact-user-locations-through-triangulation/},
  2021.
\newblock Accessed: 17-Oct-2023.

\bibitem{illegal_telegram}
{Telegram Receives Massive Fine For Failing To Report Illegal Content}.
\newblock
  \url{https://www.digitalinformationworld.com/2022/10/telegram-receives-massive-fine-for.html},
  2022.
\newblock Accessed: 17-Oct-2023.

\bibitem{news_telegram}
{Don't Use Telegram's New 'People Nearby' Feature}.
\newblock
  \url{https://lifehacker.com/dont-use-telegrams-new-people-nearby-feature-1846017886},
  2023.
\newblock Accessed: 17-Oct-2023.

\bibitem{backlinko_telegram}
{How Many People Use Telegram in 2023? 55 Telegram Stats}.
\newblock \url{https://backlinko.com/telegram-users}, 2023.
\newblock Accessed: 17-Oct-2023.

\bibitem{mapGithub}
{Telegram Nearby Map}.
\newblock \url{https://github.com/tejado/telegram-nearby-map}, 2023.
\newblock Accessed: 17-Oct-2023.

\bibitem{abu2017_eurousec}
Ruba Abu-Salma, Kat Krol, Simon Parkin, Victoria Koh, Kevin Kwan, Jazib
  Mahboob, Zahra Traboulsi, and M~Angela Sasse.
\newblock {The Security Blanket of the Chat World: An Analytic Evaluation and a
  User Study of Telegram}.
\newblock In {\em Proceedings of EuroUSEC '17}. Internet Society, 2017.

\bibitem{albrecht2022_sp}
Martin~R. Albrecht, Lenka Mareková, Kenneth~G. Paterson, and Igors Stepanovs.
\newblock {Four Attacks and a Proof for Telegram}.
\newblock In {\em 2022 IEEE Symposium on Security and Privacy (SP)}, pages
  87--106, 2022.

\bibitem{alkhulaiwi2016_pst}
Rakan Alkhulaiwi, Abdulhakim Sabur, Khalid Aldughayem, and Osama Almanna.
\newblock {Survey of secure anonymous peer to peer Instant Messaging
  protocols}.
\newblock In {\em 2016 14th Annual Conference on Privacy, Security and Trust
  (PST)}, pages 294--300. IEEE, 2016.

\bibitem{anglano201731}
Cosimo Anglano, Massimo Canonico, and Marco Guazzone.
\newblock Forensic analysis of telegram messenger on android smartphones.
\newblock {\em Digital Investigation}, 23:31--49, 2017.

\bibitem{barsocchi2020_icqict}
Paolo Barsocchi, Antonello Calabr{\`o}, Antonino Crivello, Said Daoudagh,
  Francesco Furfari, Michele Girolami, and Eda Marchetti.
\newblock {A Privacy-By-Design Architecture for Indoor Localization Systems}.
\newblock In {\em International Conference on the Quality of Information and
  Communications Technology}, pages 358--366. Springer, 2020.

\bibitem{wei2011}
Wei Chang, Jie Wu, and Chiu~C. Tan.
\newblock Friendship–based location privacy in mobile social networks.
\newblock {\em International Journal of Security and Networks}, 6(4):226--236,
  2011.

\bibitem{dargahi2017_cikm}
Arash Dargahi~Nobari, Negar Reshadatmand, and Mahmood Neshati.
\newblock {Analysis of Telegram, an instant messaging service}.
\newblock In {\em Proceedings of the 2017 ACM on Conference on Information and
  Knowledge Management}, pages 2035--2038, 2017.

\bibitem{nobari2017}
Arash Dargahi~Nobari, Negar Reshadatmand, and Mahmood Neshati.
\newblock Analysis of telegram, an instant messaging service.
\newblock In {\em Proceedings of the 2017 ACM on Conference on Information and
  Knowledge Management}, CIKM '17, page 2035–2038, New York, NY, USA, 2017.
  Association for Computing Machinery.

\bibitem{ermoshina2021telegram}
Ksenia Ermoshina and Francesca Musiani.
\newblock The telegram ban: How censorship “made in russia” faces a global
  internet.
\newblock {\em First Monday}, 26(5), 2021.

\bibitem{guhl2020safe}
Jakob Guhl and Jacob Davey.
\newblock A safe space to hate: White supremacist mobilisation on telegram.
\newblock {\em Institute for Strategic Dialogue}, 26, 2020.

\bibitem{hartle2022impact}
Frank~X Hartle, Megan Garfinkel, David O'Neil, Gaetano Scalise, Nicholas Sauer,
  and Christopher Willard.
\newblock The impact of social media geolocation on national security and law
  enforcement.
\newblock {\em Issues in Information Systems}, 23(1), 2022.

\bibitem{kargar2018censorship}
Simin Kargar and Keith McManamen.
\newblock Censorship and collateral damage: Analyzing the telegram ban in iran.
\newblock {\em Berkman Klein Center Research Publication}, (2018-4), 2018.

\bibitem{lee2017security}
Jeeun Lee, Rakyong Choi, Sungsook Kim, and Kwangjo Kim.
\newblock Security analysis of end-to-end encryption in telegram.
\newblock In {\em Simposio en Criptograf{\'\i}a Seguridad Inform{\'a}tica,
  Naha, Jap{\'o}n. Disponible en https://bit. ly/36aX3TK}, 2017.

\bibitem{huaxin2018}
Huaxin Li, Haojin Zhu, Suguo Du, Xiaohui Liang, and Xuemin Shen.
\newblock Privacy leakage of location sharing in mobile social networks:
  Attacks and defense.
\newblock {\em IEEE Transactions on Dependable and Secure Computing},
  15(4):646--660, 2018.

\bibitem{li2010}
Nan Li and Guanling Chen.
\newblock Sharing location in online social networks.
\newblock {\em IEEE Network}, 24(5):20--25, 2010.

\bibitem{ludant2023_sp}
Norbert Ludant, Pieter Robyns, and Guevara Noubir.
\newblock {From 5G Sniffing to Harvesting Leakages of Privacy-Preserving
  Messengers}.
\newblock In {\em 2023 IEEE Symposium on Security and Privacy (SP)}, pages
  3146--3161, 2023.

\bibitem{morgia2023_icws}
Massimo~La Morgia, Alessandro Mei, Alberto~Maria Mongardini, and Jie Wu.
\newblock {It’s a Trap! Detection and Analysis of Fake Channels on Telegram}.
\newblock In {\em 2023 IEEE International Conference on Web Services (ICWS)},
  pages 97--104, 2023.

\bibitem{vicente2011}
Carmen Ruiz~Vicente, Dario Freni, Claudio Bettini, and Christian~S. Jensen.
\newblock Location-related privacy in geo-social networks.
\newblock {\em IEEE Internet Computing}, 15(3):20--27, 2011.

\bibitem{rosler2018_eurosp}
Paul Rösler, Christian Mainka, and Jörg Schwenk.
\newblock {More is Less: On the End-to-End Security of Group Chats in Signal,
  WhatsApp, and Threema}.
\newblock In {\em 2018 IEEE European Symposium on Security and Privacy
  (EuroS\&P)}, pages 415--429, 2018.

\bibitem{Sutikno2016WhatsAppVA}
Tole Sutikno, Lina Handayani, Deris Stiawan, Munawar~Agus Riyadi, and Imam
  Much~Ibnu Subroto.
\newblock Whatsapp, viber and telegram: which is the best for instant
  messaging?
\newblock 2016.

\bibitem{kokevs2017}
Tom\'{a}\v{s} Su\v{s}\'{a}nka and Josef Koke\v{s}.
\newblock Security analysis of the telegram im.
\newblock ROOTS, New York, NY, USA, 2017. Association for Computing Machinery.

\bibitem{vaziripour2018_usec}
Elham Vaziripour, Justin Wu, Reza Farahbakhsh, Kent Seamons, Mark O’Neill,
  and Daniel Zappala.
\newblock {A Survey of the Privacy Preferences and Practices of Iranian Users
  of Telegram}.
\newblock In {\em Workshop on Usable Security (USEC)}, volume~1, 2018.

\bibitem{walther2021us}
Samantha Walther and Andrew McCoy.
\newblock Us extremism on telegram.
\newblock {\em Perspectives on Terrorism}, 15(2):100--124, 2021.

\bibitem{wernke2014classification}
Marius Wernke, Pavel Skvortsov, Frank D{\"u}rr, and Kurt Rothermel.
\newblock A classification of location privacy attacks and approaches.
\newblock {\em Personal and ubiquitous computing}, 18:163--175, 2014.

\end{thebibliography}

\end{document}